\documentclass{article}

\usepackage{arxiv}
\usepackage[utf8]{inputenc} 
\usepackage[T1]{fontenc}    
\usepackage{hyperref}       
\usepackage{url}            
\usepackage{booktabs}       
\usepackage{amsfonts}       
\usepackage{nicefrac}       
\usepackage{microtype}      
\usepackage{lipsum}		
\usepackage{graphicx}
\usepackage{natbib}
\usepackage{doi}

\usepackage{longtable}

\begin{document}

\title{Opinion Change or Differential Turnout: Changing Opinions on the Austin Police Department in a Budget Feedback Process}



\author{ \href{https://orcid.org/0000-0003-1936-638X}{\includegraphics[scale=0.06]{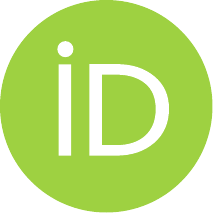}\hspace{1mm}Lodewijk L. Gelauff}\\
	Management Science and Engineering\\
	Stanford University\\
	Stanford, CA 94301 \\
	\texttt{lodewijk@stanford.edu} \\
	\And
    \hspace{1mm}Ashish Goel\\
	Management Science and Engineering\\
	Stanford University\\
	Stanford, CA 94301 \\
	\texttt{ashishg@stanford.edu} \\
}
\maketitle

\begin{abstract}
  In 2020 the tragic murder of George Floyd at the hands of law enforcement ignited and intensified nationwide protests, demanding changes in police funding and allocation. This happened during a budgeting feedback exercise where residents of Austin, Texas were invited to share opinions on the budgets of various city service areas, including the Police Department, on an online platform designed by our team. Daily responses increased by a hundredfold and responses registered after the ``exogenous shock'' overwhelmingly advocated for reducing police funding.
  
  This opinion shift far exceeded what we observed in 14 other Participatory Budgeting elections on our Participatory Budgeting Platform, and can't be explained by shifts in the respondent demographics. Analysis of the results from an Austin budgetary feedback exercise in 2021 and a follow-up survey indicates that the opinion shift from 2020 persisted, with the opinion gap on police funding widening. We conclude that there was an actual change of opinion regarding police funding.
  
  This study not only sheds light on the enduring impact of the 2020 events and protests on public opinion, but also showcases the value of analysis of clustered opinions as a tool in the evaluation toolkit of survey organizers. 
\end{abstract}



\keywords{participatory budgeting, natural experiment, opinion change, police funding, clustering, minority representation}


\section{Introduction}
\label{sec:introduction}

The year 2020 brought about significant turbulance in the relationship between U.S. cities and their residents. It was marked by a tense nationwide political climate \citep{Abramowitz2019}, the COVID pandemic altering the dynamics of resident engagement with city governments, and several high-profile, fatal encounters between law enforcement and Black individuals that resulted in widespread protest and demands for change. The police budget and organization came under varying degrees of scrutiny and there was also an increased interest in health and human services  \citep{krieger_enough_2020, maynard_police_2020}. 

One of the most prominent incidents in this period was the tragic killing of George Floyd by law enforcement officers in Minneapolis on May 25, which was captured on shocking video footage that ignited protests across the United States \citep{Barrie2020}. This happened during an ongoing online budget feedback exercise that we conducted in partnership with the City of Austin. In this exercise, Austin residents were asked to provide feedback on both the revenue and the expenditure side of the city budget, including the allocation of funds to the Austin Police Department (APD). Remarkably, we observed a hundredfold increase in daily responses to the budget feedback exercise and a discernable shift in how respondents allocated funds to different city functions. In line with econometrics literature, we will refer to the immediate aftermath of George Floyd's murder as an ``exogenous shock'', effectively transforming our exercise into a natural experiment. In the subsequent year, we partnered once again with the City of Austin to conduct a scaled-down version of the budget feedback exercise, accompanied by a follow-up survey asking participants directly whether (and how) their opinion of police funding changed over the preceding two years.

We produced summary reports for the City of Austin detailing the outcomes of these exercises, including aggregated budgets. In 2020, in the wake of the protests and citizen engagement, the City announced a range of measures that had an impact on the budget of the Austin Police Department, including a cut of 11 million (2.5\%) \citep{flores_newly_2020}, which was in line with the aggregated budget in our report. In addition to the civic impact of our work, we believe that the nature of the exogenous shock, the swift change in the modal response on the platform with respect to police funding, the clear importance of providing equitable voices to different segments of the city, and the fact that we repeated a similar exercise a year later make our data and analysis interesting from a research perspective as well. We focus on these research implications rather than a normative or policy-oriented analysis of the desirability of cuts in police funding.

\subsection{Summary of Research Contributions and Findings}
We detail the budget feedback exercise's design and timeline in Section~\ref{sec:process}, modeled after Participatory Budgeting elections~\citep{rudas_participatory_2021, Goel2019}.  Participants answered detailed questions about revenue items, and reallocated funds among city functions, ensuring a balanced budget.
In Section~\ref{sec:outcomes}, we highlight the most salient aspects of our data. The unedited daily quantitative reponses, stripped of demographic and identifying information, have been publicly released\footnote{Refer to Section~\ref{sec:conclusions} for data specifics}. Given the unique nature of this natural experiment, we believe that this data is independently valuable, both to researchers interested in broader social choice issues of equitable voice and equitable participation, and to those specifically interested in sentiments towards police funding before and after the George Floyd murder.

In Section~\ref{sec:analysis} we analyze the data along multiple dimensions. In 2020, after the exogenous shock of George Floyd's murder, participant opinion unmistakenly swung towards substantial police funding reductions. While the underlying issues with defunding the police were inextricably linked to racial justice, support for decreasing police funding was notably higher among White than among Black respondents. Demographic shifts do not adequately explana the dramatic opinion change after the shock. 

We employ cluster analysis, encompassing both quantitative and qualitative responses, uncovering a more nuanced picture of the participants who wanted to reduce police funding; we argue that such nuanced cluster analysis can guide more informed and representative policy decisions. 
We also perform a comparitive analysis with Participatory Budgeting elections, revealing that our data has a swing well beyond what we observe in those elections.  

The pivotal question is whether the 2020 opinion shift (especially regarding the police funding) after the shock resulted from opinion change, a change in the composition of the participant population, and whether the opinion change (if any) was lasting. The 2021 budgetary feedback exercise and a follow-up survey provide useful insights. Demographics in the 2021 exercise resembled pre-shock 2020 demographics, yet support for further police funding rediction persisted. Further, our follow-up survey in 2021 suggests that there was a change in public opinion, and the differential outcomes before and after the shock in 2020 were not just the result of a temporary change in the participant turnout. The cluster analysis also allows us to compare the results of the follow-up budgeting survey in 2021 with the results from 2020 at a high level, and we find the cluster compositions to be qualitatively different.

We acknowledge that our study's scope is limited to one city and budgeting exercise, and advise against hasty generalizations. 
Further, the dimensions along which we are concerned about equitable voice may differ across countries, and the optimum design of the budget interface may also depend on access to technology in the populations being surveyed. It is important to note that this was not a randomized study, and the participants were self-selected. Also, the results of the 2021 exercise were eliciting public opinion relative to a new baseline.

While we emphasize data and analysis in this paper, our methodological contributions inform future design and evaluation of such feedback exercises. Our work highlights that residents can meaningfully engage with complex budget balancing processes, and other cities are already adopting similar approaches\footnote{For example, https://abalancingact.com/ uses some similar elements and is being used by several U.S. cities.}.
While a reduction in police funding was the most striking aspect of the aggregated budget, we highlight that there was substantive and useful information in how participants chose to reallocate police funding. The design choice that each participant had to submit a response subject to budget constraints was consequential. 

The fact that these clear and robust clusters exist is an important finding for future high-dimensional civic feedback processes, and our approach to computing these clusters (normalizing along each axis and then running k-means) can serve as a useful starting point. 
In Section~\ref{sec:clustering budgeting} we show that voting data from similar Participatory Budgeting elections can indeed often be clustered, and that the progression of the fraction of votes per cluster over the course of the election, can provide helpful insights. 

Our study raises normative issues about representation and equity of voice in civic feedback processes. Unline binding elections (which generally conform to one person one vote), these exercises offer advisory results, allowing for reweighting to ensure equitable voice, but the appropriate approach remains uncertain. The broader question on reweighting along demographic lines, or employing clustering-based methods to address ``opinion minorities'' deserves careful exploration. Our work serves as one data point in this ongoing and more comprehensive exploration of how to design civic feedback processes that are robust against disparities in the representation of demographic and opinion minorities.

\subsection{Related Work}
Citizen participation in government is a well-established and multivarious practice, mandated since the Equal Opportunity Act of 1964 in various federal initiatives in the United States\citep{callahan_elements_2007}. \cite{langton_citizen_1978} classified citizen participation into the categories of electoral participation, citizen action (grass roots activism), citizen involvement (``initiated and controlled by the government''), and obligatory participation. We will focus here on citizen involvement\footnote{We will assume these terms to include all residents.}, also known as public participation (expanding the scope beyond citizens). The anticipated impacts could be a way to better meet the needs of residents, an opportunity to build consensus and to improve public trust of the decision making process \citep{wang_assessing_2001, callahan_elements_2007}. Wang's 2001 survey found that 46\% of U.S. cities' chief administrative officers reported citizen involvement in the budgeting process~\citep{wang_assessing_2001}. Calls to involve residents, stakeholders or citizens in the budgeting process are nothing new \citep{ebdon_citizen_2006} and \cite{ebdon_beyond_2002} found that a third of the cities in her sample actively sought citizen input, and a fifth did so on the entire operating budget -- although (nearly) all cities use the traditional method of public hearings to some extent. Callahan concludes however that ``public hearings do little more than inform the public'' and that direct citizen participation (using a stricter definition) is not widely adopted by public administrators \citep{callahan_citizen_2007}. Ebdon and Franklin \citep{ebdon_citizen_2006} identified public meetings, focus groups, budget simulations, citizen advisory committees and citizen surveys as mechanisms that can be engaged, all with their own advantages and constraints. The 2020 (and 2021) Austin budget feedback exercise that we're discussing in this paper was designed on the intersection between a budget simulation and a resident survey -- taking advantage of an educational element but also as a way to determine and report preferences.

Perhaps one of the more empowering ways to get residents engaged in parts of the city budget has been budget allocation through Participatory Budgeting, a process that has been well described in the literature (e.g. \cite{ganuza_conflicts_2016, rudas_participatory_2021}). Since the first process under this name is often considered to have been in Porto Alegre, Brazil, more than 7,000 cities across the world have run a process under that can be considered part of the Participatory Budgeting family \citep{dias_next_2018}. The process exists in many different variations and with various definitions, ranging from a grassroots process in the style of Porto Alegre where residents take active control of a significant portion of the city budget to a more modest approach where residents are informed and consulted on public finances, or get to make a choice between some well-defined projects \citep{sintomer_participatory_2008}. The history of how PB spread and was adapted in various cities has been extensively described in the literature \citep{bartocci_journey_2022}. 

In recent years, it has been increasingly common for organizers of participatory budgeting processes to give residents the option to express their preferences through an online tool. A number of software platforms has been made available for cities to use, both commercially and non-commercially. This includes, but is far from limited to, the Stanford Participatory Budgeting Platform\footnote{https://pbstanford.org} \citep{Goel2019}, Consul\footnote{https://consulproject.org} \citep{arana-catania_citizen_2021, pina_decide_2022} and Decidim \citep{serramia_optimising_2019}. In this publication, we will use data from the Stanford Participatory Budgeting Platform, a platform that is primarily used in North America. 

Within the context of Participatory Budgeting, where the organizer wants to select a number of projects given budgeting constraints, there is a range of different voting methods available to choose from. A voting method in this context consists on both elicitation (the ballot used to collect opinions from the voter) and aggregation (how an allocation among budget items is determined based on the votes). It would be well beyond the scope of this paper to discuss the different voting methods, but detailed surveys of relevant social choice publications are available in \cite{rey_computational_2023, rudas_participatory_2021}. The knapsack voting method is most relevant in this paper. Knapsack elicitation is a form of (constrained) approval voting (where a voter can select any projects that they approve of) with an added budget constraint: the sum of the costs of the selected projects can not exceed the budget of that election. Under this model, voters naturally include the cost of a project as in their considerations. The strategic properties and effectiveness of this elicitation method are discussed in \cite{Goel2019}. 
 
In the United States, goals to organize PB can often include equity \citep{lerner_budgeting_2020}, and analysis in New York found that such a process can indeed have a positive effect on the likelihood that people from traditionally underrepresented groups participate in regular elections \citep{johnson_testing_2021}. 
In an election setting, it would generally be unacceptable to reweight responses from different sub-populations to arrive at a more representative outcome. However, this method could be applied in feedback exercises such as ours to help the decision maker visualize potential alternative outcomes. A lot of progress has been made in identifying techniques to address the challenge of interpreting results of a non-representative or respondent-driven survey \citep{mercer_for_2018, Gelman2016,wejnert_web-based_2008}. These methods assume access to instrumental variables that capture the under representation. Another approach is to design the survey from the start to be more representative, be it by using weighted advertising methods \citep{gelauff_advertising_2020}, adjusted survey design \citep{berg_inclusivity_2020} or by adjusted sortition (for minipublics) \citep{flanigan_fair_2021}. 
Random or balanced population samples form the basis of deliberative polls \citep{fishkin_when_2009} and can also be used in deliberative budgeting processes, requiring a smaller sample but more engagement, allowing for in-depth and more informed discussion and opinions \citep{ackley_community_2021, wilson_deliberative_2020}. Increasingly, cities are using more complex voting methods or feedback processes and with the further availability of online technology, novel methods such as the exercise described in this study can be implemented at more reasonable cost.

In recent years, commentators around the world have been surprised by sudden shifts in opinion polls, or by the fact that opinion polls did not reflect the eventual outcome during an election or referendum. While these polls of national elections have been shown to generally perform rather well \citep{jennings_election_2018}, smaller polls and referendums like the Brexit referendum have resulted in surprises. In general these surprises can be attributed to 5 factors: Actual change of opinion (`late swing'), differential turnout, swing voters behaving differently from determined voters, misstatement of opinion by polled citizens or non-representative samples \citep{mellon_missing_2017, kennedy_evaluation_2018}. The change in responses after the exogenous shock in our exercise raises similar questions about the cause and nature of the change.

Our work contributes to recent literature examining the effects of police killings of Black citizens (such as Ahmaud Arbery, Breonna Taylor and George Floyd) and subsequent protests in the context of a worldwide pandemic \citep{christian_background_2022}. Previous research identified a shift in sentiment towards Black individuals on Twitter after all three killings, with the strongest shift after George Floyd's murder \citep{nguyen_progress_2021}.
Especially relevant are two studies that compare survey data before and after the murder on George Floyd and found unprecedented increase in anger and sadness levels among the US population beyond Minnesota \citep{eichstaedt_emotional_2021}, an increase in distress on police brutality among young people \citep{howard_young_2022} and a shift in police favorability, where previous killings led to limited effects only \citep{reny_opinion-mobilizing_2021}. This suggests that the intensity of protests and national outrage after the murder on George Floyd was fundamentally different from cases before - whether it is due to the nature of the murder or the combination with the ongoing pandemic. 

\section{The Budget Feedback Exercises: Timeline and Design}
\label{sec:process}
\subsection{Timeline}
\begin{figure*}[ht]
\small
    \centering
    \begin{minipage}{\linewidth}
        \centering
        \includegraphics[width=0.90\linewidth, alt={Histogram of responses per day}]{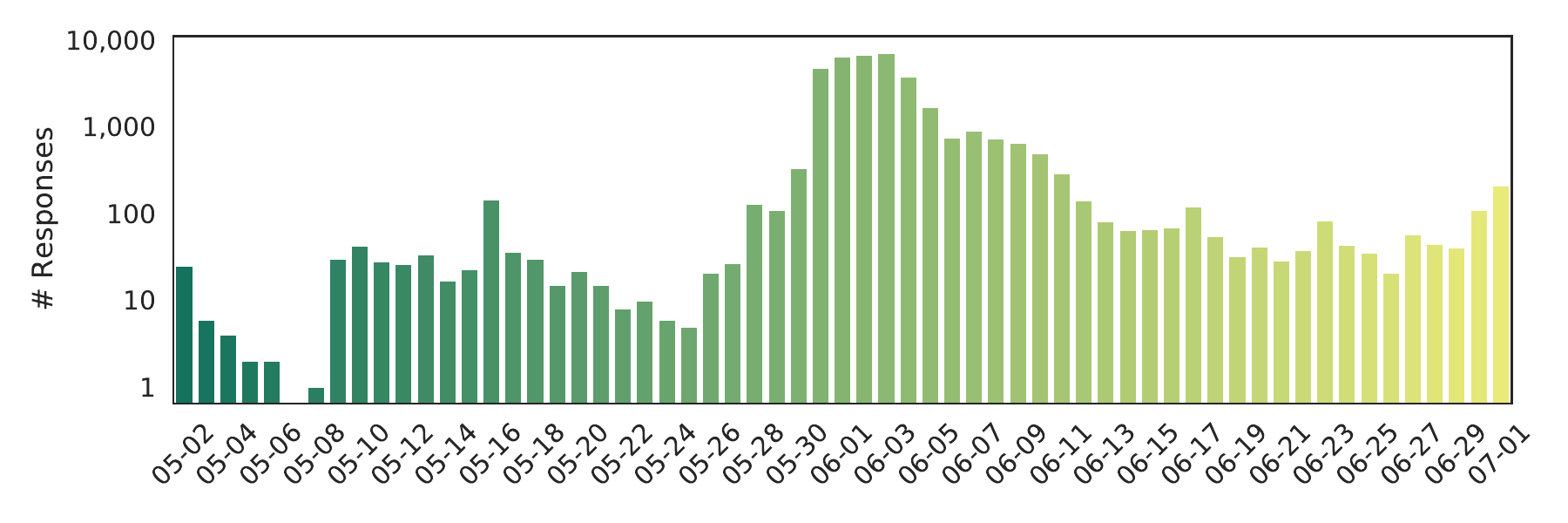}
        \setlength{\abovecaptionskip}{0pt}
        \caption{Responses per day (2020) on a logarithmic scale}
        \label{fig:date hist 2020}
    \end{minipage}
\end{figure*}
In late 2019, the City of Austin budget office found itself constrained in its revenue sources while simultaneously facing rapidly increasing expenses due to the rising cost of living in the city. The General Fund revenue, as reported by the City, amounted to \$1.1 billion in FY2020 with the property tax (49\%), sales tax (23\%) and utility transfers (15\%) being the primary revenue sources. The City had limited control over the size of these revenue sources due to a state-enforced ceiling on the revenue growth from property tax. 
This meant that some tough budgetary choices would likely have to be made. In early 2020, a more nuanced feedback exercise was collaboratively designed with the authors to gather resident input into this complex multi-dimensional issue. The exercise design was minimally affected by the emerging Covid pandemic, other than some adjustments in the introduction. However, its importance increased due to reduced opportunities for the City to collect in-person feedback from residents as the pancemic unfolded. 

The exercise was launched as a custom-made website on May 1, 2020 and initially it was planned to run for one month. This timeline was extended to two months due to the limited availability of offline opportunities for residents to provide input on the city budget and the surge in responses toward the end of the original month. The City promoted the website through its usual communication channels, including social media, news letters and traditional media, with an additional effort to ensure responses from traditionally underrepresented areas and populations. The research team had no direct involvement in the advertising of the exercise. 

On May 25, George Floyd was killed in Minneapolis, an event that sparked widespread outrage and led to a surge in protests and criticism of police funding across the country. This heightened citizen engagement had a notable impact on the response rate (Fig.~\ref{fig:date hist 2020}). Up until May 25, daily responses were (often much) less than 100, but they rapidly increased to more than 1,000 responses per day between May 31 and June 6. Following June 10, the daily responses remained elevated, averaging 98 per day. For analysis and presentation, we will divide the responses in three non-overlapping segments: May 1-29 (segment 1), immediately after the exogenous shock May 29 - June 6 (segment 2), and June 6-30 (segment 3). 

A technical report of the outcomes was prepared and a first draft was shared with City leadership mid June and a final report published on July 21 with an aggregated operating budget from the responses that reduced police funding by 3\% \citep{Chen2020}. In the wake of the protests and citizen engagement, the City announced a range of measures that impacted the budget of the Austin Police Department, including a cut of 11 million (2.5\%) \citep{flores_newly_2020}, which was in line with the reported outcomes of the exercise. The City Council eventually decided on a larger redistribution of the safety budget \citep{van_oudenaren_did_2020}, which was mostly reverted in 2021 after new state legislation penalized cities that cut their police budget \citep{McGlinchy2021}. 
In 2021, the collaboration was continued with a redesigned budget feedback exercise with 1237 respondents \citep{Gelauff2021}. 

\subsection{Design}
\label{sec:Design}
\begin{figure*}[ht]
    \centering
    \begin{minipage}{0.48 \linewidth}
        \centering
        \small
        \includegraphics[width=0.75\linewidth, alt={Screenshot with brief explanation of revenue side of the budget, and two example fee categories.}]{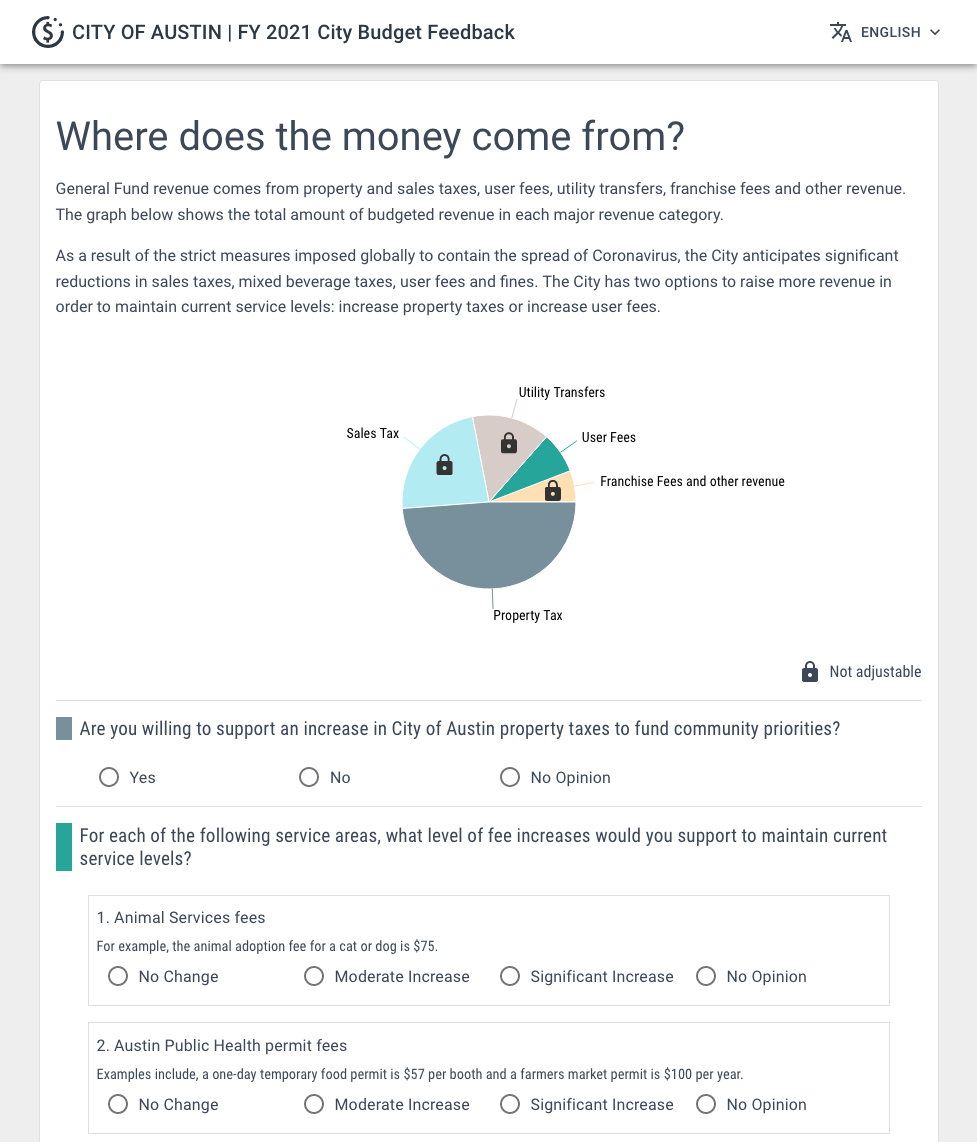}
        \setlength{\abovecaptionskip}{0pt}
        \caption{Design of revenue portion of the 2020 exercise.}
        \label{fig:design revenue 2020}
    \end{minipage}\hfill
    \begin{minipage}{0.45\linewidth}
        \centering
        \small
        \includegraphics[width=0.75\linewidth, alt={Screenshot with brief explanation of expenditure side of the budget and two example expense categories.}]{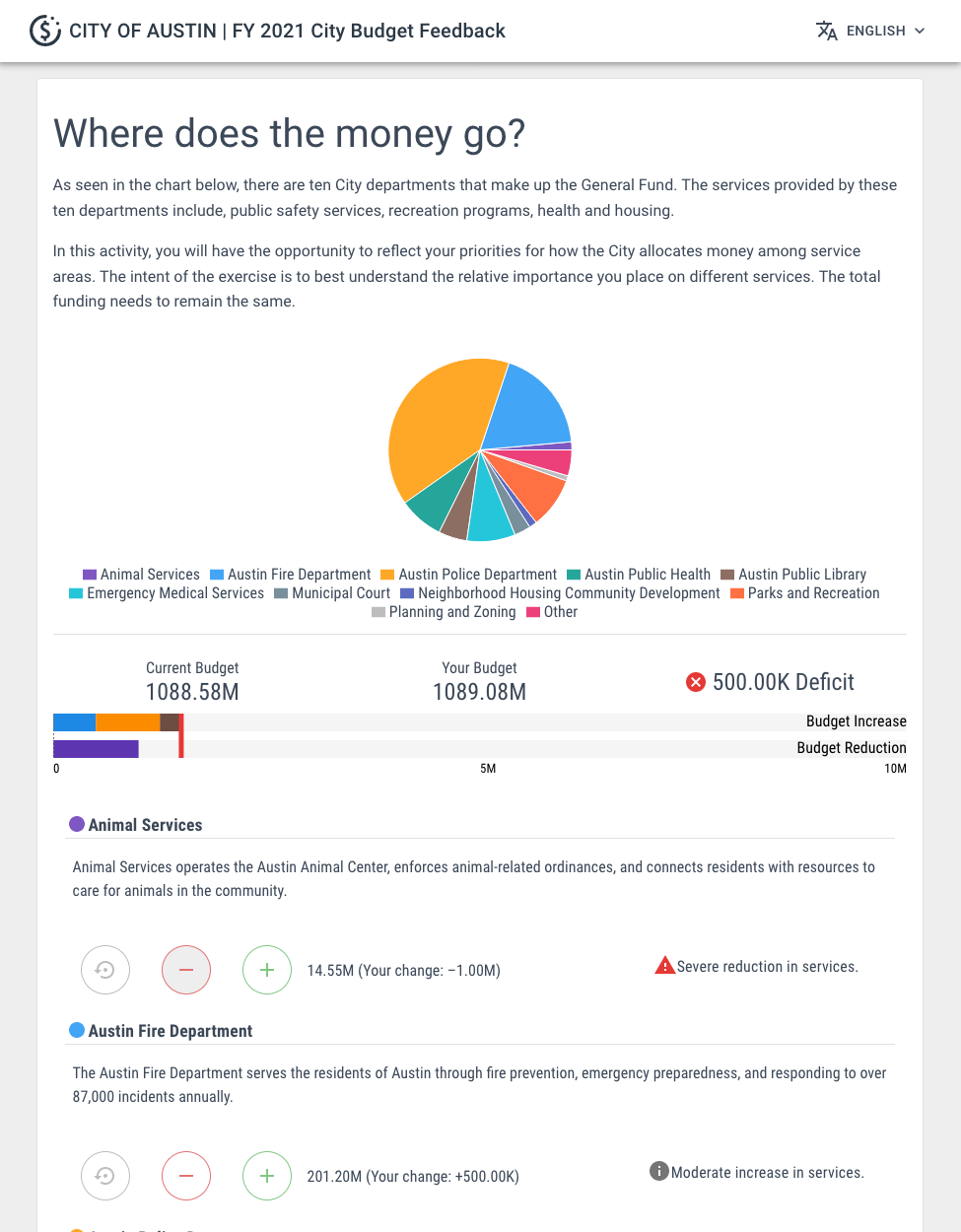}
        \setlength{\abovecaptionskip}{0pt}
        \caption{Design of expenditure portion of the 2020 exercise.}
        \label{fig:design expenditure 2020}
    \end{minipage}
\end{figure*}
The 2020 edition was designed to get input on two questions: 
\begin{enumerate}
    \item If the City would have to increase revenue, what would be the support to increase property tax or service fees;
    \item If residents would be able to redistribute the budget between departments (zero sum), how would they do so. 
\end{enumerate}

The workflow was designed with three components: revenue, expenditure and demographic. The authentication was designed to be a very low threshold for participation, with self-certification that the respondent lived in the City of Austin and use of reCaptcha. In the revenue portion, residents were asked what their support was for increasing the property tax, and for which service area they would support a moderate or significant increase of fees in 9 service areas (e.g. aquatic fees and golf fees). Because these service areas are often diverse, it was not possible to provide quantitative questions for these fees and they were posed qualitatively (see Fig. \ref{fig:design revenue 2020}).

Next, respondents were presented with the current distribution of the General Fund across 11 city service areas (e.g. Austin Police Department and Emergency Medical Services) and asked to redistribute the budget between services in \$250,000 increments (see Fig.~\ref{fig:design expenditure 2020}). To ensure a realistic scenario, any respondent could not reduce the budget of any department by more than 5\%, and in order to submit this section of the exercise, the respondent had to arrive at the same total. This budgeting under constraints provides a more holistic view of what the respondents are actually interested in. While this design was not originally set up as a referendum on the police budget, we did anticipate that the APD was likely a divisive budget item. This approach would provide a contextualized and consistent way of asking about the entire budget, even if the budget of the APD was of most interest to respondents. 

Finally, respondents were presented with a set of demographic questions and a few open-ended opportunities to provide input or feedback. 

In 2021, the design of the expenditure section was changed from a redistribution of budget to a five-point scale to only allow residents to indicate per service area whether they would support a significant or moderate decrease, no change or a moderate or significant increase of its budget. The service areas also reflected the new service areas of the city (most notably the Public Safety Support, which was split off from the APD) and the exercise was due to its new design this time hosted on a popular off-the-shelf survey website. 

In 2021, we also invited respondents to volunteer for a follow-up survey, where we asked them about their opinion on the APD budget and to choose between three aggregated sets of preferences. We took the responses of 2021 and clustered the responses to get 3 scenarios for revenue, and 3 scenarios for expenditures from their respective centroids (removing the items with general agreement between cluster centroids).\footnote{The scenarios are available in the Appendix Section \ref{app:followup 2021}} We presented these scenarios in randomized triplets and asked participants to provide their preferred order. 

We explained the changes made to the APD budget in the ongoing financial year as explained on the City website\footnote{\url{https://www.austintexas.gov/news/austin-city-council-approves-fiscal-year-2020-2021-budget}}, and asked whether they agreed with these changes in APD funding. We also asked whether their idea about the ideal size of the Police force has changed over the previous 1-2 years, and asked an open question as to what the most important event was from the past 1-2 years that changed their opinion on the APD budget. 
The exercise designs and the follow-up survey were approved by the Stanford University Institutional Review Board.

\section{Feedback Exercise Outcomes}
\label{sec:outcomes}

We will briefly present the most salient outcomes of the two feedback exercises and the 2021 follow-up survey, before moving on to further analysis.

\subsection{Outcomes 2020}
\begin{figure*}[ht]
    \centering
    \begin{minipage}{0.48 \linewidth}
        \centering
        \small
        \label{fig:outcomes revenue 2020}
        \includegraphics[width=\linewidth, alt={Table with level of support per cell, with background color indicating size of support}]{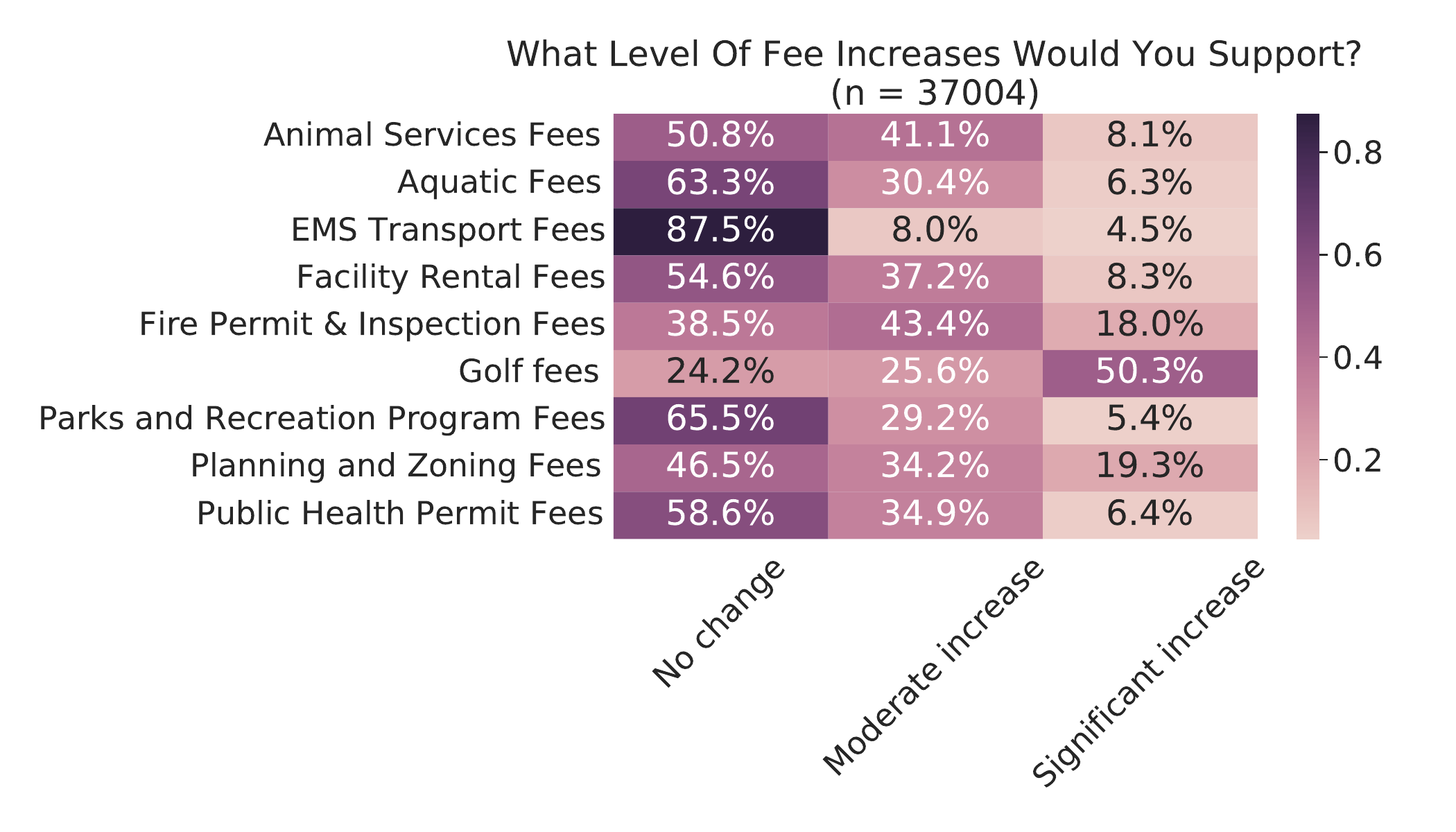}
        \setlength{\abovecaptionskip}{0pt}
        \caption{Responses per fee category, what level of fee increases  (revenue) would be supported. Background color indicates size of support. Significant means 3\% or more.}
    \end{minipage}\hfill%
    \begin{minipage}{0.48\linewidth}
        \centering
        \small
        \label{fig:outcomes expenditure 2020}
        \includegraphics[width=\linewidth, alt={Table with level of support per cell, with background color indicating size of support}]{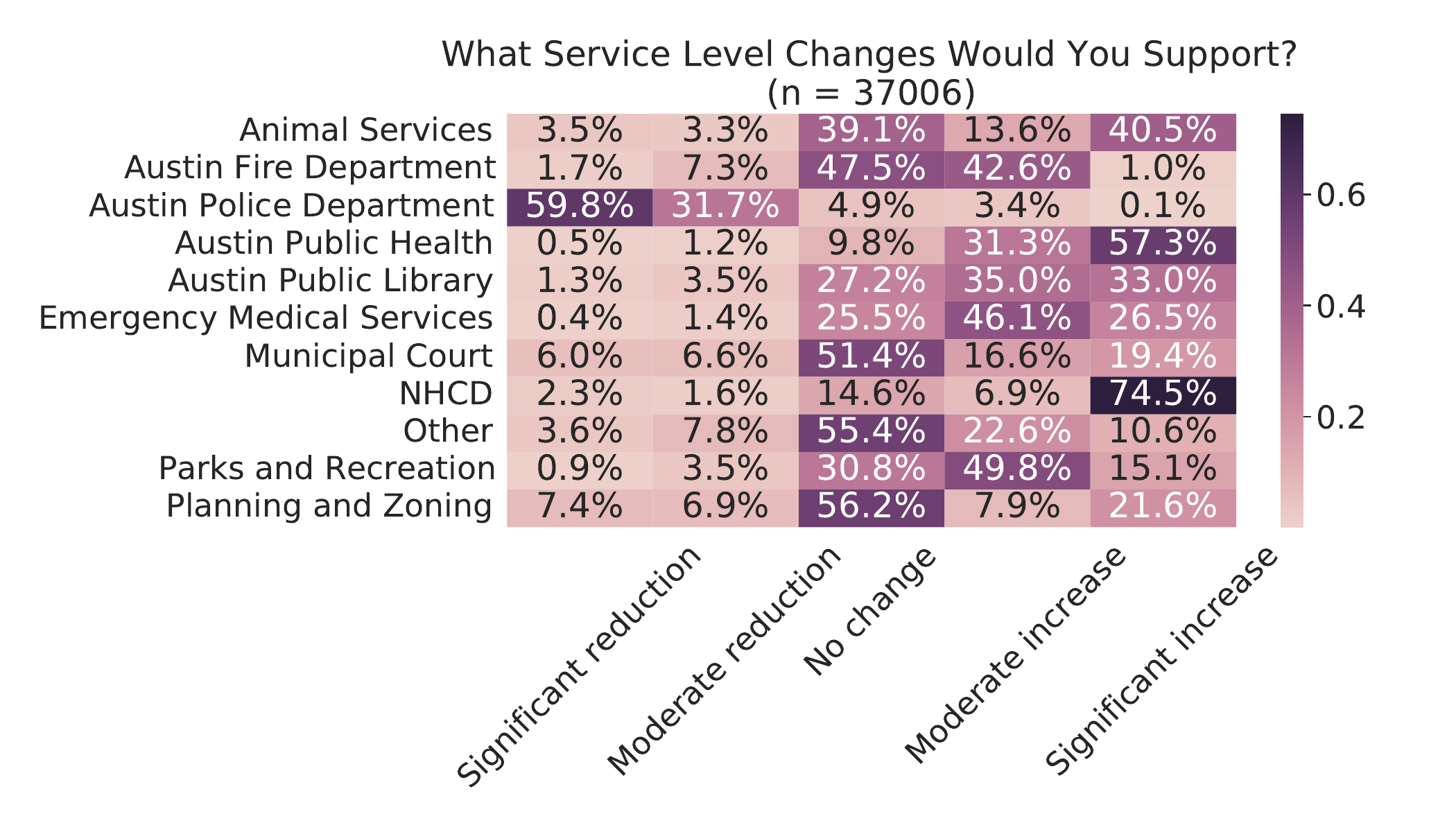}
        \setlength{\abovecaptionskip}{0pt}
        \caption{Responses per service area, what level of expenditure increases would be supported. Background color indicates size of support. Significant means 3\% or more.}
    \end{minipage}
\end{figure*}

The 2020 exercise received 37,006 responses up until July 1. Most notable demographic deviations from the American Community Survey~(ACS)~2018~\citep{MCDC2020} or census data were an over/under representation of some districts, the proportion of respondents 18-34 years old (0.73 respondents, 0.30 ACS) and of respondents 55+ (0.035 respondents, 0.20 ACS).\footnote{Tables in Appendix section~\ref{app:Demographics} show the demographic distribution (gender, district, age, race/ethnicity, home ownership and income) of respondents. }

The aggregated results of the 2020 responses have previously been released in a technical report to the City management~\citep{Chen2020}.\footnote{Tables with aggregated results are provided in the Appendix section~\ref{app:outcomes 2020}} In summary:
\begin{itemize}
    \item 50.1\% of respondents (n=37,006) was willing to support an increase in property taxes (35.4\% not willing to support, 14.5\% no opinion).
    \item Out of the 9 service areas, there was in three areas a majority to support fee increases: golf fees, fire permit \& inspection fees and planning and zoning fees. (n=37,004)
    \item For the expenditures (n=37,006), there was 59.8\% support to reduce the budget of the police department with more than 3\%, and 91.5\% support for some reduction. This budget had then to be allocated to other service areas, resulting in increased budgets for almost all other departments. These budget preferences were then aggregated with knapsack aggregation which essentially finds a multi-dimensional median of all the submitted budgets subject to the budget-balance constraint~\citep{Goel2019}. The aggregated results, presented in Table~\ref{tab:expenditure 2020 knapsack}, include a change of $-3$\% for the Police Department, with the bulk of these funds being redirected to the Public Health, Emergency Medical Services and NHCD (Neighborhood Housing and Community Development).
\end{itemize}

\subsection{Outcomes 2021}
The 2021 exercise received 1525 valid responses.\footnote{Demographic distribution of respondents is available in the Appendix section~\ref{app:Demographics}} The most visible deviations from the ACS 2019~\citep{MCDC2021} were the over/under representation of some districts, the proportion of respondents 18-24 years old (0.052 respondents, 0.10 ACS), the proportion of Latinx/Hispanic (0.10 respondents, 0.33 ACS) and White respondents (0.79 respondents, 0.49 ACS) and the proportion of renters (0.31 respondents, 0.47 ACS). These proportions are much more similar to the respondents from before the exogenous shock, than after. 

The aggregated results of the 2021 responses have previously been released in a technical report to the City management~\citep{Gelauff2021}.\footnote{Tables with aggregated results are provided in the Appendix section \ref{app:outcomes 2021}} In summary:
\begin{itemize}
    \item 28\% of the respondents (n=1525) indicated a support for an increase in property taxes (65\% no support, 7\% no opinion)
    \item Only for three service areas, a majority of respondents supported an increase in service fees: Golf Fees, Fire Permit \& Inspection Fees and Facility Rental Fees. (n=1525)
    \item For the expenditures (n=1400), the questions changed compared to 2020. The proposed budget was significantly different (with ‘Public Safety Support’ split off from the Police budget) and the question no longer had an internal balance constraint. The police budget saw 33\% support for significant decrease and 28\% for significant increase. A similar split (with more support for increase) was on the opinions regarding Public Safety Support. 
\end{itemize}

\subsection{Follow-up Survey}
\begin{table*}[htp]
    \centering
    \small
    \caption{Cross tabulation ``Do you currently agree with [the APD budget] changes'' and ``In the last 1-2 years, has your idea of the ideal size of the police force changed''. $\rho = -0.37, p=0.000$}
    \label{tab:followup 2021 crosstab police}
    \begin{tabular}{p{0.20\textwidth}p{0.15\textwidth}p{0.15\textwidth}p{0.15\textwidth}p{0.15\textwidth}}
        \toprule
        Agreement with current changes: & No, the Police needed more funding, not less &  No, the change was in the right direction, but too much &  Yes, I agree &  No, the changes were in the right direction, but it was not enough  \\
        Opinion on size police force changed in the past 1-2 years  &   &   &    &    \\
        \midrule
        Yes, I now believe that the ideal size is larger    &   40 &    5 &     4 &         0   \\
        No, my opinion is about the same                    &   16 &    13 &    33&        5    \\
        Yes, I now believe that the ideal size is smaller   &   4 &     6  &    41 &       28   \\
        \bottomrule
    \end{tabular}
\end{table*} 

The 2021 follow-up survey received 204 responses, of which 163 were matched to the 2021 exercise. 40\% of the respondents (n=198) indicated that they agreed with the APD budget changes, while an additional 29\% indicated that the change was in the right direction; 31 \% indicated that the Police needed more funding, not less. The respondents that said that their ideal size is larger than before (25\%), also mostly indicated that APD needed more funding, the respondents that said their ideal size didn't change (34\%) were centered around agreement with the changes and respondents that said that ideal size is smaller than before (40\%) were split between agreement with changes, and wanting larger changes (n=196).\footnote{We refer to Table~\ref{tab:followup 2021 crosstab police} for more detail.} 

We asked respondents to rank three different aggregated revenue and expenditure scenarios in order to determine persistence of opinion. When comparing the revenue scenarios (n=127), 47\% selected scenario rev-B (moderate increase of most service fees, but not of the property tax) as their first choice, and on the expenditure side (n=135), 59\% selected scenario exp-A (significant decrease of APD budget, moderate increase for most other areas).\footnote{We refer to Table \ref{tab:followup 2021 rankings} in the Appendix for more detail.}

\section{Analysis}
\label{sec:analysis}

\subsection{Shift effect}
After observing the jump in responses (see Fig.~\ref{fig:date hist 2020}), we first established that this was not due to illegitimate responses by inspecting the open-ended responses and user agent strings, as well as IP-addresses used for submission. We observed a level of variation that is contradictory with a small number of respondents responding many times. We verified that responses were mostly from residents: in the demographic questions, 3\% entered a zip code that was not associated with the city, and 76\% of the respondents provided a valid combination of city council district number and zip code, if both questions were answered. 70\% of the responses could be mapped to Austin through IP address, and 95\% to Texas. Some imperfect mapping is to be expected, and thus gives no reason to expect that the increase could be explained by responses submitted by people who did not live or work in Austin. These signals suggest that the submissions were likely manual and primarily made by the target audience. 

News coverage of the budget feedback exercise was rather limited, but there is some evidence of social media posts getting traction. 
For example we were able to find some activity on Twitter, with the tweet with largest reach was retweeted 700 times and many tweets were identified with a smaller reach, many using a slightly different screenshot of the exercise website. This is consistent with the image that the peak was unlikely caused by a single organizer and more likely by a broader interest in the topic of the exercise once word got out about the exercise being organized on behalf of the City, grasping at an opportunity to provide a signal to the City -- even as the possibility that some organized external effort was performed to attract a specific audience to the exercise, cannot be entirely excluded.

The most visible shift in demographics is that participants in segment 2 were younger and more likely to rent their home, rather than to own it. The most eye catching opinion shift is in the expenditures\footnote{See Appendix Section~\ref{app:outcomes time} for outcomes split out by segment.}: the support for increasing the Police budget drops from 22\% to 3\% and the support for decreasing the budget increases from 43\% to 93\%, with the other departments seeing an expected support for increased budget due to the balanced expenditure requirement. At the same time, we observe a decrease in support for increased service fees across the board (e.g. support for increasing Facility Rental Fees going from 67\% to 40\%) and support for a property tax increase, increased from 41\% to 52\%. The responses in segment 3 bounce back to a limited extent in the direction of segment 1. 

There is a shift in opinions on increased service fees between segments 1 and 2. However, when service areas are sorted by their support within a time segment, the service fees at the top (most support to increase: golf fees and fire permit \& inspection fees) and bottom (least support: EMS transport fees) don't change within the segments, or between years. 
The knapsack-aggregated budget from before the exogenous shock was identical to the default budget proposed by the city, but in segment 2 and 3, this aggregated to a budget reduction for the police department of 2.99\% (\$13 million), and the budget was distributed over other departments.  

While there are clear distinctions in responses between the segments, it is not obvious whether the opinion shift is caused by a shift in turnout rates (people with one opinion, or people with another opinion), or whether societal opinion shifted. We will try to address this with the follow-up survey analysis. 

\subsection{Responses across demographics}
\label{sec:opch responses across demographics}
A first step is to inspect whether the demographics shifted in a meaningful way between segment 1 and 2. As far as meaningful shifts happened, they happened in the direction of the demographic distribution of the city, sometimes overshooting the ACS. Gender and race ratios saw with limited shifts, but some demographic shifts were notable: 
\begin{itemize}
    \item Individuals with a household income of less than \$35,000 went from 12\% to 23\% (ACS: 23\%)
    \item Renters went from 36\% to 66\% (ACS: 50\%)
    \item Participants 18-24 year old went from 3\% to 30\% (ACS: 11\%), 25-34 year old went from 28\% to 46\% (ACS: 23\%)
    \item Some shifts in districts (suburban district~8 went from 10\% to 5\%, while inner city district~9 went from 8\% to 15\%)
\end{itemize}

Participants in segment 2 were younger, had a lower income and were more likely to rent than people who participated in segment 1. In segment 3, these demographics return more to levels of segment 1, but not quite. In 2021, the participation of young people, renters and lowest income mostly return to that of segment 1. 

Splitting out the responses by demographics gives some useful insights. For example, we observe that the support for property tax increases varies by age (18-24 years old: 54\% support, 75+: 18\%) and home ownership (home owners 45\%, renters 54\%), that the support for increasing service fees varies with age (younger groups support increases less across the board) and much less with home ownership or race. Also on the expenditure side, age is a meaningful demographic with more support to reduce police funding among young respondents, while there seems little connection with race and some connection with home ownership. 

To verify whether the shift in opinions before and after the exogenous shock could be explained by a different demographic turnout, we aggregated the responses in all three time segments, reweighted by age, home ownership or race/ethnicity to match distributions in the ACS. We still see similar meaningful shifts across the board in these adjusted aggregates.\footnote{As age is the biggest difference, we made age-adjusted numbers available in Tables \ref{tab:revenue 2020-1 age} -- \ref{tab:expenditure 2020-3 age} in the Appendix}

\subsection{Cluster analysis}
\label{sec:cluster analysis}
\begin{figure*}[ht]
    \centering
    \begin{minipage}{0.48 \linewidth}
        \centering
        \small
        \includegraphics[width=\linewidth, alt={Horizontal bar chart displaying the distribution of the cluster centroids}]{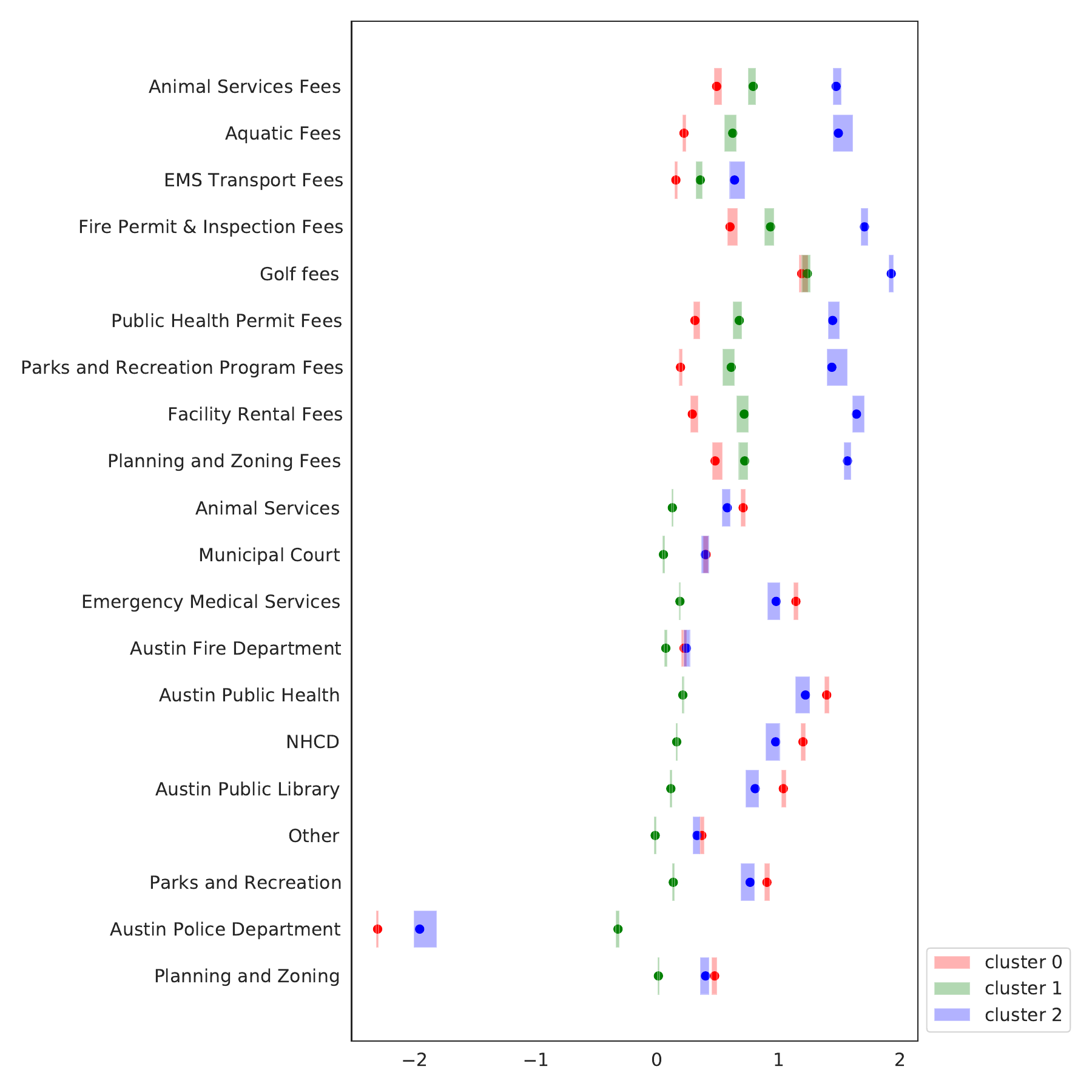}
        \setlength{\abovecaptionskip}{0pt}
        \caption{3 cluster-centroids (2020) with their normalized budget shift per service area, for 3 clusters with 0.95 confidence interval}
        \label{fig:bootstrap 2020 3clusters}
    \end{minipage}\hfill
    \begin{minipage}{0.48\linewidth}
        \centering
        \small
        \includegraphics[width=\linewidth, alt={Horizontal bar chart displaying the distribution of the cluster centroids}]{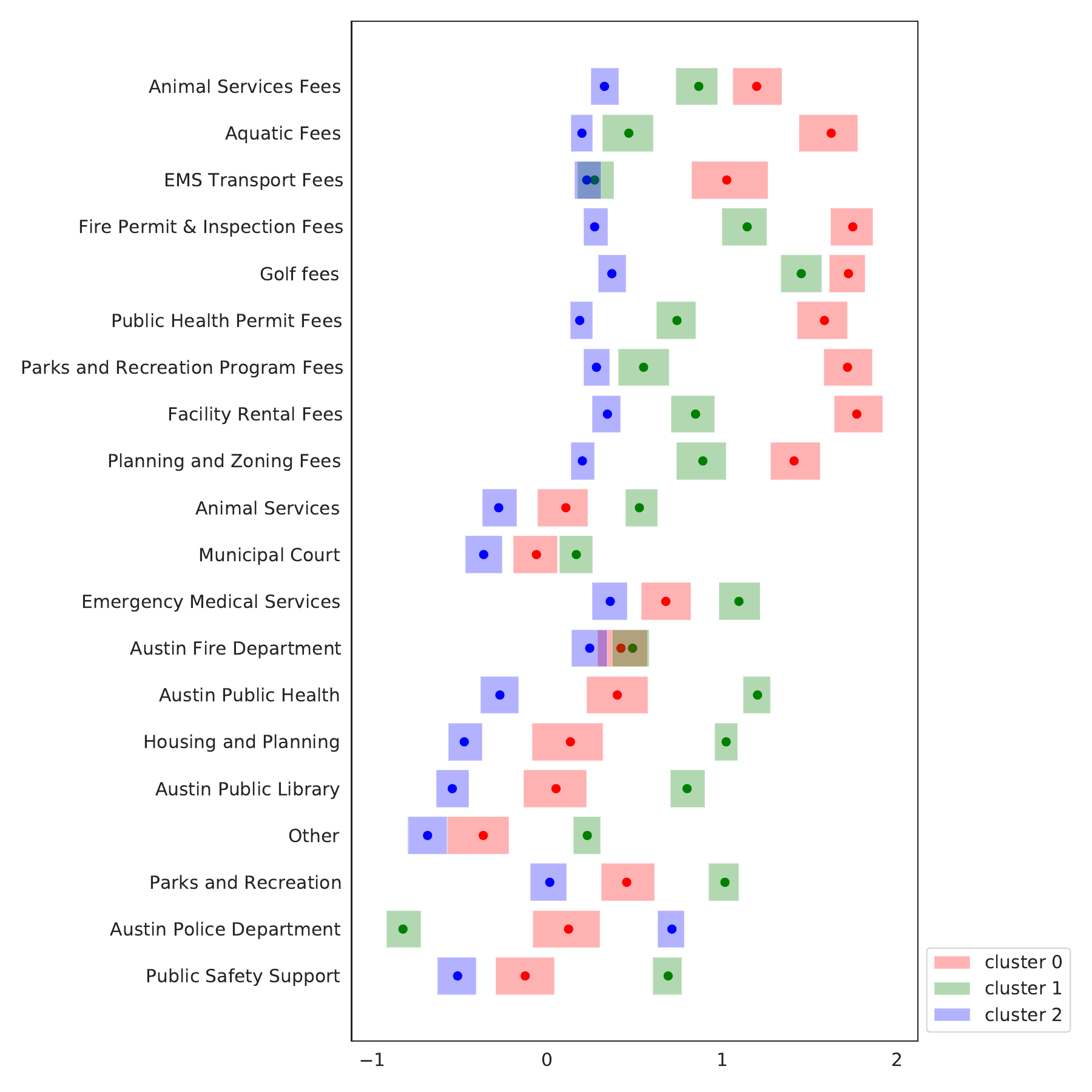}
        \setlength{\abovecaptionskip}{0pt}
        \caption{3 cluster-centroids (2021) with their normalized budget shift per service area, for 3 clusters with 0.95 confidence interval}       
        \label{fig:bootstrap 2021 3clusters}
    \end{minipage}
\end{figure*}

Because we could not satisfactorily explain the effects of the exogenous shock with demographics, we clustered the responses as a different approach to analyze the results. We converted the responses to the individual service-area based revenue and expenditure questions to numerical responses and normalized them by dividing by the standard deviation for each sub question. We used KMeans clustering since it is one of the most common methods to cluster high-dimensional data~\citep{wu_top_2008}. We configured the algorithm to find three clusters in our data (98.6\% average accuracy: same cluster label assigned to a response after resampling and reclustering) and added the labels to our dataframe. We repeated the same for 2021 data (96.7\% average accuracy). We tried 2, 3 and 4 clusters, and for this dataset the setting with 3 clusters provides the sharpest insights and picture. In Fig.~\ref{fig:bootstrap 2020 3clusters} and~\ref{fig:bootstrap 2021 3clusters} we show the mean of the normalized scores for each of the clusters and its centroids' 95\% confidence interval for 3 clusters. The centroids of both 2020 and 2021 clusters are robust to re-sampling and re-clustering. A clustering with 2 clusters provides a wider confidence interval for 2020 data, and the centroid confidence intervals of 4 clusters are less well separated.\footnote{For reference we have included the equivalent figures for 2 and 4 clusters in the Appendix.} 

\begin{figure*}[ht]
    \centering
    \begin{minipage}{0.48 \linewidth}
        \centering
        \small
        \includegraphics[width=\linewidth, alt={graph with proportion of respondents from the three clusters, in different colors. Before the first dashed line, the green cluster Cluster 1 dominates, after the second dashed line the clusters are similar. }]{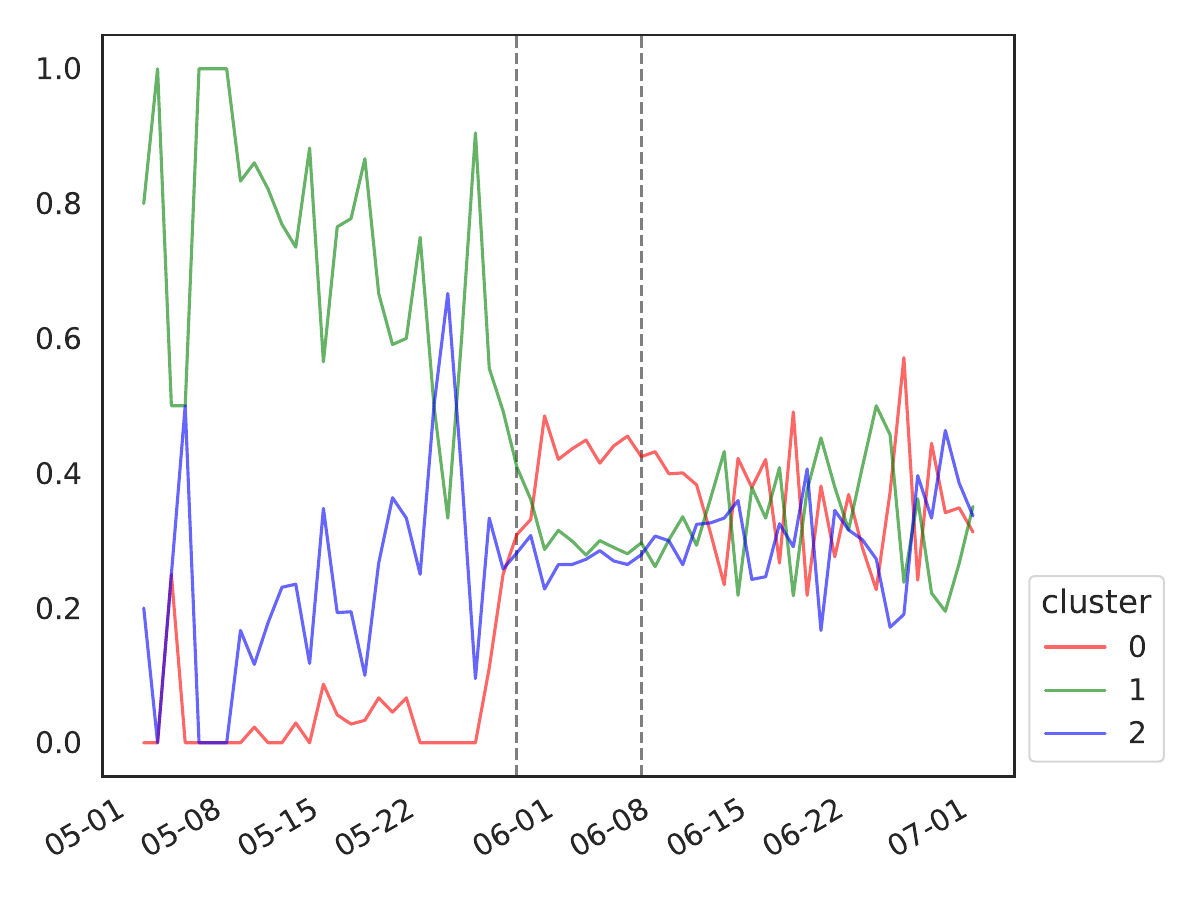}
        \setlength{\abovecaptionskip}{0pt}
        \caption{Proportion of respondents per cluster, per day for 2020. The vertical dashed lines show the boundary between time segments.}
        \label{fig:date ratio clustersplit 2020}
    \end{minipage}\hfill
    \begin{minipage}{0.48 \linewidth}
        \centering
        \small
        \includegraphics[width=\linewidth, alt={graph displaying the progression of votes per cluster over time. Cluster 2020-1 is more than 2 standard deviations over represented at the start compared to expected fractions.}]{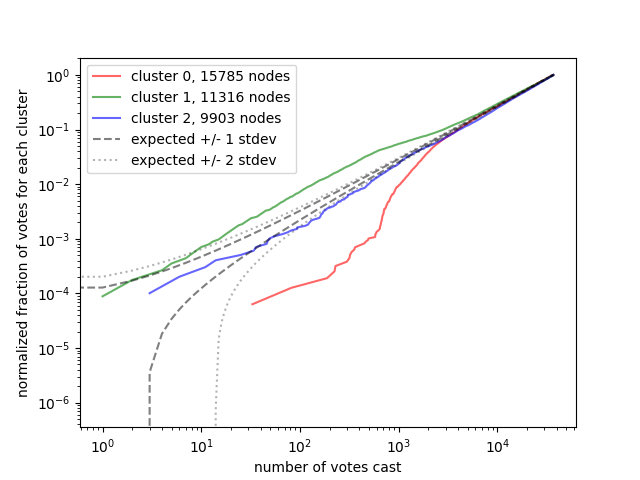}
        \setlength{\abovecaptionskip}{0pt}
        \caption{Progression of responses from each cluster for Austin 2020. The colored lines show the cumulative proportion of responses submitted per cluster. The dashed and dotted lines show respectively 1 and 2 standard deviations from the expected fraction.}
        \label{fig:progression austin 2020}
    \end{minipage}
\end{figure*}

The proportions of the different clusters over time (see Fig.~\ref{fig:date ratio clustersplit 2020}) show that cluster 2020-1 dominates among early participants, and cluster 2020-0 only appears just before the exogenous shock and being somewhat more represented immediately after the exogenous shock, and a mostly even mix of the three clusters after. 
In Fig.~\ref{fig:progression austin 2020} we show the fraction of votes per cluster that has been cast at each time in the process. The fraction of votes from cluster 2020-1 is more than 2 standard deviations higher than the expected fraction at the start of the process, while the fraction of votes from clusters 2020-0 and 2020-2 are more than 2 standard deviations lower than expected. We see a similar pattern when we analyze the independently clustered qualitative responses in Section~\ref{sec:open-ended responses}. 

When we inspect the mean opinions per cluster, we see three profiles of voters arise in 2020 data:
\begin{itemize}
    \item Cluster 2020-0 respondents are least supportive of increasing service fees, are most in favor of reducing police funding, and also increasing the funding of other departments. These respondents are generally younger and more likely to rent their home.
    \item Cluster 2020-1 respondents are least supporting of decreasing police funding or increasing the tax rate, and are more moderate on service fees. These respondents are generally older.
    \item Cluster 2020-2 respondents are supportive of meaningful reduction of police funding, but also supportive of increasing service fees and property tax. 
\end{itemize}

We should remind ourselves that the setup changed in 2021: the status quo is now that the police funding has already been partially diverted in a safety fund, but there is also no longer a requirement to balance the budget. We now find these cluster profiles:
\begin{itemize}
    \item Cluster 2021-0 respondents are most supportive of increasing service fees, and are moderate on changing expenditure or increasing property tax. These respondents are more likely to be older.
    \item Cluster 2021-1 respondents are most supportive of further reducing the police funding and increasing the budget of other departments, and have different opinions on increasing service fees, depending on the service area. They are most supportive of increasing the property tax. These respondents are more likely to be younger, female and to rent their home. 
    \item Cluster 2021-2 respondents are most supportive of increasing police funding, and reducing the budget of other departments (or increase them less), while also being least supportive of increasing service fees and property tax. These respondents are more likely to be older and to be male. 
\end{itemize}

There are some parallels between these two sets of cluster characteristics. Clusters 2020-1 and 2021-2 have in common the relative support for police funding, the relatively low support for funding other departments (in 2020, this might have been caused by the question setup, in 2021 this was not the case) and the relatively low support for property tax increase. If we however look at the cluster in 2021 that was least supportive of police funding (2021-1), we see some parallel with both 2020-0 and 2020-2: relatively high support for funding other departments, relatively high support for property tax increase. However, when we look at the fee increases on the revenue side of the budget, this parallel breaks down: in 2020 one of the clusters was most supportive of increasing fees, while the other was least supportive. In 2021, this seems to have merged together in a single cluster that supports increasing some fees much more than others. This is in line with the picture that the turnout of people who supported reducing police funding in a meaningful way was much higher in 2020 than in 2021.

At a high level, we can conclude that we find robust respondent profiles that may be helpful for the decision maker. While there are some parallels between 2020 and 2021 clusters, it is due to changes in survey setup and society unrealistic to expect that the cluster profiles will match from year to year.

\subsection{Follow-up Survey}

In the follow-up survey, we asked whether respondents currently agreed with the changes made to the police budget and split out the responses by cluster in Table \ref{tab:followup police currently}. These preferences are consistent with the cluster centroids in the 2021 clustering. 

We also asked respondents how their opinion of the ideal size of the police force changed over the past 1-2 years, and find a correlation with their agreement to the implemented APD budget changes (See Table~\ref{tab:followup 2021 crosstab police}). Respondents who want more funding for the police, also believe that their ideal Police force is now larger than before, and vice versa. 
In other words: most of the respondents that have the most extreme opinions with regards to police funding, have developed or reinforced that opinion over the past 1-2 years. 

\begin{table}[htp]
    \centering
    \small
    \caption{Count of preferred scenario by respondents in follow-up survey, split by 2021 cluster. The clusters are significant for scenario preference with $p = [0.04, 0.00, 0.00]$, using a $\chi^2$ test comparing to the overall (unclustered) response distribution.}
    \label{tab:followup preferred scenario}
    \begin{tabular}{lcccccc}
    \toprule
    & \multicolumn{3}{c}{Revenue scenarios}     
    & \multicolumn{3}{c}{Expenditure scenarios} \\
    \cmidrule(lr){2-4}                  
    \cmidrule(lr){5-7}
     &  rev-A &  rev-B &  rev-C &  exp-A &      exp-B &  exp-C \\
    cluster label & & & & & & \\
    \midrule
    0   &       9 &         19 &       2 &      15 &        12 &     8 \\
    1   &       5 &         28 &       24 &      48 &       4 &      1 \\
    2   &       17 &        5 &        3 &       7 &        14 &     6 \\
    Total &     31 &        52 &       29 &     70 &        30 &    15 \\
    \bottomrule
    \end{tabular}
\end{table}

\begin{table*}[htp]
    \centering
    \small
    \caption{Count of follow-up survey respondents by agreement with police budget change and cluster.
    Clusters are significant with $p = [0.04, 0.00, 0.00]$, using $\chi^2$ test comparing to the overall (unclustered) response distribution}
    \label{tab:followup police currently}
    \begin{tabular}{lp{0.15\linewidth}p{0.15\linewidth}p{0.15\linewidth}p{0.15\linewidth}}
    \toprule
     cluster& No, the Police needed more funding, not less & No, the change was in the right direction, but too much & Yes, I agree & No, the changes were in the right direction, but it was not enough \\
    \midrule
    0   &   22 &    8 &     14  & 4  \\
    1   &   2 &     9 &     45  & 26 \\
    2   &   26 &    5 &     6   & 1  \\
    Total & 50 &    22 &    65  & 31 \\
    \bottomrule
    \end{tabular}
\end{table*}

In Table \ref{tab:followup preferred scenario} we show for each 2021 cluster what the count of preferred scenarios\footnote{A detailed description of the content of each scenario is available in Appendix Section \ref{app:followup 2021}.} was. We observe that preferences are in line with the cluster centroids in the original 2021 clustering. We also analyzed the data from the question about agreement with the change in police budget (Table \ref{tab:followup police currently}), which is consistent with the cluster centroids from the 2021 feedback exercise. 

\section{Clusters of qualitative and participatory budgeting data}
\label{sec:clustering budgeting}

\subsection{Open-ended responses}
\label{sec:open-ended responses}
\begin{figure*}[ht]
    \centering
    \begin{minipage}{0.31 \linewidth}
        \centering
        \small
        \includegraphics[width=\linewidth]{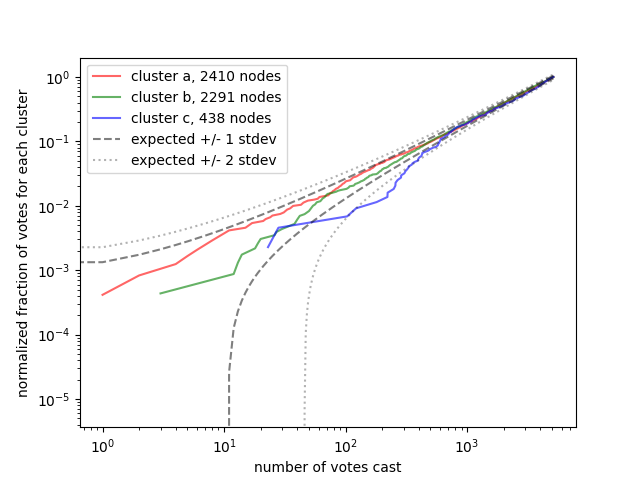}
        \setlength{\abovecaptionskip}{0pt}
        \caption{Progression of responses from each cluster on open-ended revenue elaboration.}
        \label{fig:progression votes revenue}
    \end{minipage}\hfill
    \begin{minipage}{0.31 \linewidth}
        \centering
        \small
        \includegraphics[width=\linewidth]{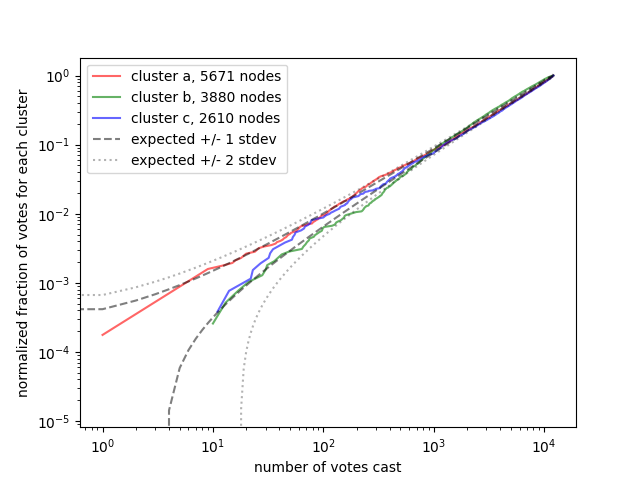}
        \setlength{\abovecaptionskip}{0pt}
        \caption{Progression of responses from each cluster on open-ended expenditure elaboration.}
        \label{fig:progression votes expenditure}
    \end{minipage}\hfill
    \begin{minipage}{0.31 \linewidth}
        \centering
        \small
        \includegraphics[width=\linewidth]{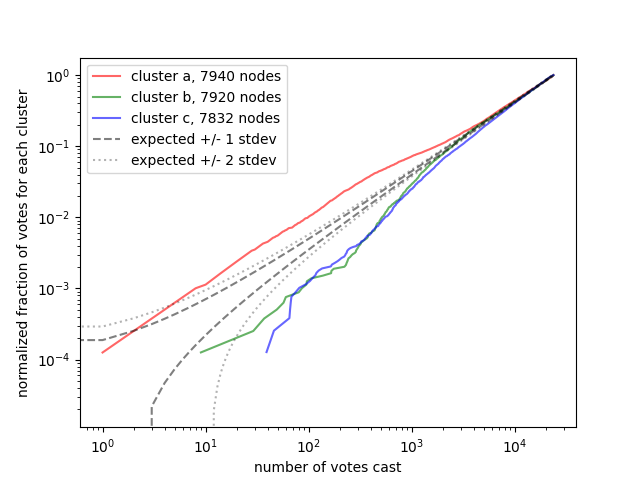}
        \setlength{\abovecaptionskip}{0pt}
        \caption{Progression of responses from each cluster on open-ended reason for participation.}
        \label{fig:progression votes reason}
    \end{minipage}
\end{figure*}
The Austin 2020 survey also contained three open-ended questions that respondents could consider\footnote{This data is not included in the publicly available dataset, to protect respondents' privacy.}:
\begin{enumerate}
    \item revenue elaboration: At the bottom of the revenue section, a field "please explain why" was included, allowing respondents to explain why they gave those responses. 5,039 non-empty responses were entered.
    \item expenditure elaboration: At the bottom of the expenditure section, a field "please explain why" was included, allowing respondents to explain why they chose that budget distribution. 12,162 non-empty responses were entered.
    \item reason for participation: In the demographic survey, respondents were asked "What was the most important reason you decided to participate?". 23,693 non-empty responses were entered.
\end{enumerate}

Analyzing these responses is not trivial, and many approaches are possible. In line with the rest of the paper, we clustered these responses again into three clusters, for each of the questions. We broadly followed the suggested path in \cite{castillo_how_2021}: We cleaned the open-ended responses, removed stopwords and tokenized the responses. We then created a model on these responses with Word2Vec from Gensim \citep{rehurek_lrec} with vectors of size 100, and clustered with minibatch k-means clustering with 3 clusters. This results in 3 clusters for each of the open-ended questions with silhouette scores of respectively 0.26, 0.16 and 0.21. 

When we match these new clusters to the labels of the quantitative clusters we discuss in the rest of the paper, it is interesting to observe that the quantitative labels 0 and 2 (both in favor of reducing police funding, but with different opinions on changes to the city's revenue) have a distinctively different distribution over the qualitative revenue-clusters, while quantitative labels 0 and 1 (with the strongest disagreement on police funding) have a distinctively different distribution over the qualitative reason-clusters. 

When the progressions of the responses are plotted per qualitative cluster for each of the three questions (Fig.~\ref{fig:progression votes revenue}, \ref{fig:progression votes expenditure} and~\ref{fig:progression votes reason}) we observe that also clusters based on the open-ended responses show a clear shock effect, but mostly so when we look at the responses to the question what the most important reason was to participate. This suggests that the respondents had a significantly different reason to participate in the exercise in the earlier part of the process, than later in the process. The fact that the qualitative clusters see the same shock, but still don't entirely overlap with the quantitative clusters, shows that there is a potential to offer additional insights to the quantitative clusters. The technique of clustering responses based on their content is a promising direction for future research, both on quantitative and qualitative data.

\subsection{Comparison with Participatory Budgeting Elections}
\label{sec:comparison with PB}
In the analysis in Section~\ref{sec:cluster analysis} we make use of opinion clusters generated from k-means clustering on the quantitative opinion data. While this turns out to be a useful approach in this specific data set, it does beg the question whether determining opinion clusters is a useful approach in budgeting opinion data in general, and how common such a sharp cluster representation discontinuity is in budgeting exercises. 

In Participatory Budgeting, respondents are asked to consider a set of projects and submit their preferences as to which projects should be funded. On the Stanford Participatory Budgeting Platform (described by Goel et al. in \cite{Goel2019}), voting methods that are commonly used include \textit{K-approval} (select up to $K$ projects), \textit{K-ranking} (select up to $K$ projects and rank them in order of preference) and \textit{Knapsack} voting (select projects up to a total budget $B$). In many of these processes, while the formal decision power remains with the decision maker that organizes the process, the explicit or implied understanding is that the results of the election would be adhered to. Independently from the comparison, it is useful to consider shocks to voting patterns in these processes: the turnout is often rather small (across 39 elections in North America, it was reported to be 2.6\% on average \citep{hagelskamp_public_2016}) and in order to advance equity and diversity among their population, many organizers go out of their way to recruit participants through public outreach. It could be expected that when some of these activities are particularly successful, they may result in a differential turnout and opinion swing. In fact, we see little evidence of such opinion swings to the extent as what we observed in Austin, indicating that the outreach is likely not biased along opinion clusters. 

We have a previously cleaned and anonymized data set available from more than 100 Participatory Budgeting elections on the Stanford Participatory Budgeting (PB) Platform that used knapsack voting, which we can compare to  the Austin 2020 data in terms of their voting patterns.\footnote{This data set is part of a paper that is currently under review. We will provide a link to that publication and/or a public repository.} On the PB platform, the process from the perspective of the voter is as follows: the process is advertised by the organizer of the election (e.g. a city), and when the voter arrives at the election landing page, they are informed about the procedures. They are then invited to authenticate themselves, the specifics of which depends on the settings chosen by the organizer. The voter can then fill out the official ballot and submit their preferences. In some elections, a secondary voting method (otherwise with the same ballot content), is then presented as an optional research ballot that the voter can skip. If the voting method is knapsack, the voter is presented with all available projects, their descriptions and costs. They can then select all the projects that they approve of, and as they select a project, the costs of that project fills up the budget bar at the top of the page. Voters can only select further projects, if they fit in their remaining budget. For some projects, the organizer may have allowed partial approval, but in most cases the voter can only approve the project, or not. 

In this data set a rigorous cleaning and anonymization process was followed: only voters that used either a single-use password or sms-authentication are included, empty votes were not included and for each election the 10\% fastest and 10\% slowest voters were removed from the data set to preserve voter privacy and improve data quality. We only include elections where with at least 100 knapsack votes in this analysis. While we include votes both when knapsack is the primary voting method or the secondary voting method, all ballots are from actual voters, on actual projects during an actual election. 

For each of these elections, we take the project costs and its votes. For each vote, we have the voter ID and the amount allocated by the voter to each project. We normalize these amounts to the portion of the project cost that they approved (usually this is either 0 or 1) resulting in a vector for each voter. We then use the method described by Tibshirani et al. \citep{tibshirani_estimating_2001} to find the optimal number of clusters, which we report in Appendix Table \ref{app:tab:optimal pb clusters}. For the 46 knapsack elections that we have left, the optimal number of clusters varies from 1 to 7. This confirms that it is not uncommon to get a similar number of clusters as we observed in the Austin 2020 data (out of the 46 elections in the set, 11 had 4 clusters, 9 had 3 clusters and 6 had 2 clusters.). 

\begin{figure*}[ht]
    \centering
    \begin{minipage}{0.48 \linewidth}
        \centering
        \small
        \includegraphics[width=\linewidth]{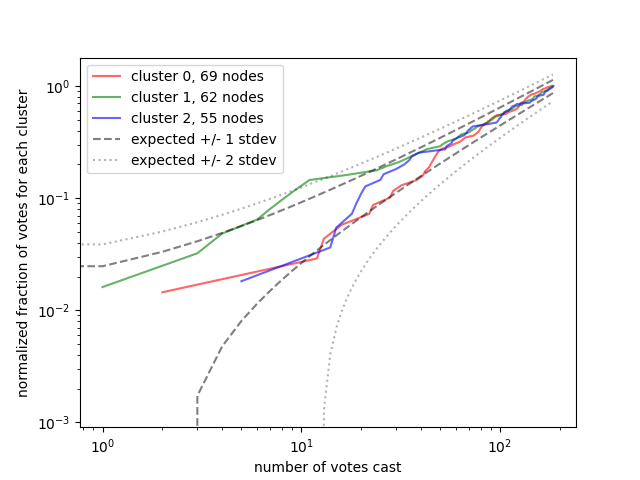}
        \setlength{\abovecaptionskip}{0pt}
        \caption{Progression of votes from each cluster for PB election 103.}
        \label{fig:progression votes 103}
    \end{minipage}\hfill
    \begin{minipage}{0.48 \linewidth}
        \centering
        \small
        \includegraphics[width=\linewidth]{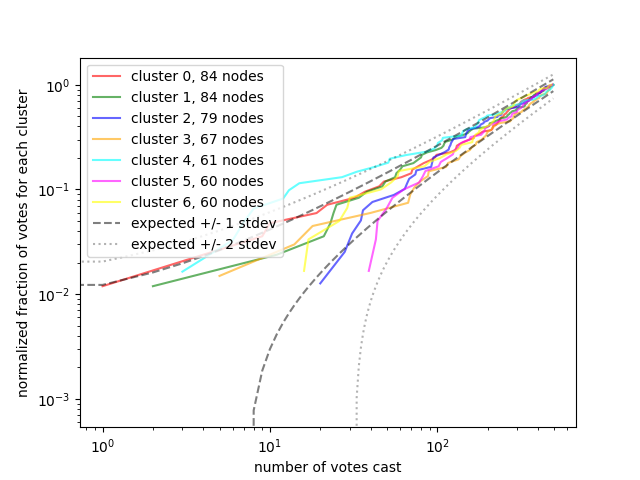}
        \setlength{\abovecaptionskip}{0pt}
        \caption{Progression of votes from each cluster for PB election 130.}
        \label{fig:progression votes 130}
    \end{minipage}
\end{figure*}

\begin{figure*}[ht]
    \centering
    \begin{minipage}{0.48 \linewidth}
        \centering
        \small
        \includegraphics[width=\linewidth]{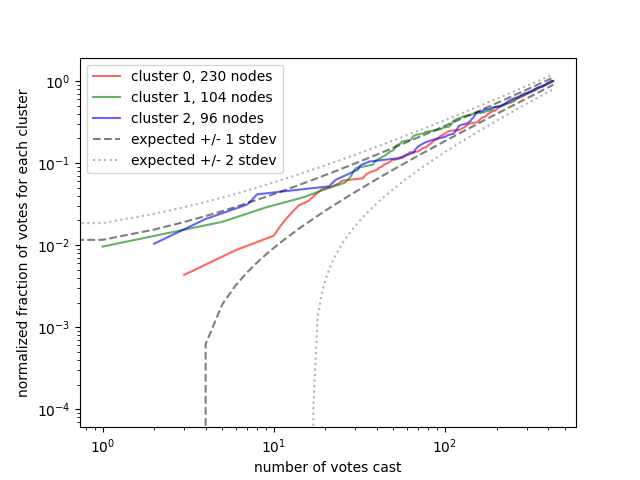}
        \setlength{\abovecaptionskip}{0pt}
        \caption{Progression of votes from each cluster for PB election 174.}
        \label{fig:progression votes 174}
    \end{minipage}\hfill
    \begin{minipage}{0.48 \linewidth}
        \centering
        \small
        \includegraphics[width=\linewidth]{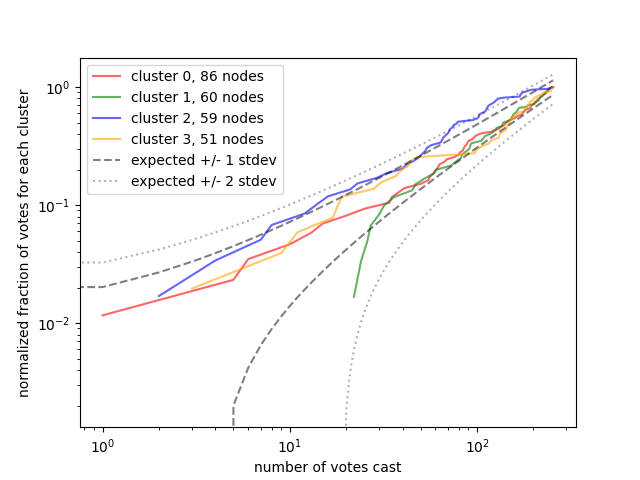}
        \setlength{\abovecaptionskip}{0pt}
        \caption{Progression of votes from each cluster for PB election 246.}
        \label{fig:progression votes 246}
    \end{minipage}
\end{figure*}

In order to understand how common the observed shift of opinions is among other budgeting processes, we analyze the shift of cluster ratios over time during these participatory budgeting elections, and compare this with the Austin 2020 data. For the elections that have an optimal cluster size of at least 3 and where the size of the smallest cluster is at least 50 votes, we then plotted the fraction of votes in each cluster that has been cast at any point in the sequence of votes over the course of the election. A few examples are displayed in Fig.~\ref{fig:progression votes 103}, \ref{fig:progression votes 130}, \ref{fig:progression votes 174} and \ref{fig:progression votes 246} and the complete set for 14 remaining elections is available in Appendix Section~\ref{app:sec:pb clustering}. We observe that in those cases, the cluster progression stays usually within 1 standard deviation of the expected value, and only gets in a few cases close to 2 standard deviations from the expected value. 

We provided the same type of figure for the Austin 2020 data\footnote{We used the published anonymized data. The order of the votes was randomized within the day. This does not meaningfully change the figure.} in Fig.~\ref{fig:progression austin 2020}, and we observe that the cumulative proportions for the first few thousand votes are well outside 2 standard deviations from the expected value. This suggests that the shift in opinion we observed in the Austin 2020 data is not typical in processes of this type. 

\section{Conclusions and Future Work}
\label{sec:conclusions}
It will come as no surprise for people who follow the politics of Austin, that the budget allocation for the Police Department is a divisive issue. This was already the case before the exogenous shock with 26\% support to increase and 43\% support to decrease the same budget item; this is even more the case in 2021 with 41\% supporting an increase (28\% a significant increase) and 45\% supporting a decrease (33\% a significant decrease) of the Police budget. After the budget changes by the City Council in 2020, the analysis of the 2021 follow-up survey shows that a large majority (70\%) could agree with the direction of these reforms, even if they didn't necessarily agree on the extent. 

We observed that Cluster 2020-1 (the only cluster not in favor of reducing police funding) dominated the responses before the exogenous shock, while after the shock the three clusters were much more balanced. This suggests that either respondents from clusters 2020-0 and 2020-2 did barely participate before the exogenous shock, or that there was an actual shift in opinion. 
In our comparison with 14 Participatory Budgeting elections, we observe that the progression of votes per cluster over time was indeed very different in the Austin data than we observe in other processes. 

We also observed that the 2021 demographics of respondents mostly returned back to those of segment 1 in 2020 and recall that the status quo had changed: the police budget had been reduced significantly, and been partially redirected to a Safety Fund. However, when we cluster opinions in 2021, we find a cluster that wanted to further reduce police funding, while clusters with that opinion were barely present in 2020 segment 1 (only 12\% of respondents in 2020 segment 1 wanted to reduce the police budget significantly). This suggests a lasting shift in opinion with regards to police funding.

We compared the 2021 cluster assignments with the responses in the follow-up survey, and conclude that the clusters capture persistent opinions. The follow-up survey also finds that in 2020/2021 the respondents that believe the police needs more funding, believe the ideal size of the Police to be larger than before, and the respondents that believe further reductions were warranted, have developed their views in the opposite direction. While there seems to have been a strong shift to reducing police funding in 2020, in 2021 the opinion gap seems to be widening.

We observed a correlation in the follow-up survey between the opinion at that time of respondents with regards to the changes in police funding, and how they indicate that their opinion on the ideal size of the police force has changed over the past 1-2 years. This suggests that the opinion gap on the police budget is widening or reinforcing. 

Police reform is a complex issue, and depends on more than just the size of the police budget. In order to really understand this kind of opinion change, it would be beneficial to dig deeper with a qualitative study that would be able to identify motivations. Further nuance could also be captured by different exercise designs, such as more detailed outlines of consequences of budget choices, or a deliberation through a mini-public. 

While we have uncovered indications that there was an opinion shift on police funding, a more conclusive and generalizable approach is to track a panel of respondents over the course of several years. Not only would this generate valuable insights on how these opinions change, it would also indicate how stable opinion clusters are over time. 

The analysis of cluster representation over time in participatory budgeting elections also demonstrates the potential of opinion clustering for evaluators for those processes. If strong shifts in submitted votes per cluster can be matched with outreach activities, they can be indicative of how successful they were in recruiting previously underrepresented opinions. At the same time, if a suspicious peak of responses coincides with a boost for a cluster of opinions well beyond what could reasonably be expected, this could be indicative of process tampering and warrant further analysis.

We have previously published a data set of responses to both feedback exercises (2020 and 2021) and the follow-up survey \citep{gelauff_austin_2022}. We rounded the timestamps to days, randomized the order within the days and selected all subjective closed questions for this data set. We believe this data may prove useful in future research on exogenous shocks. 

\section*{Acknowledgements}
The authors acknowledge the collaboration with the City of Austin and the insights that they have contributed as well as the residents that provided the feedback at the foundation of this study. We thank Yiling Chen for her contributions to the exercise design and implementation. We also appreciate the helpful discussions with and feedback from Nikhil Garg, Ravi Sojitra, Daniel Kharitonov, Mohak Goyal and anonymous reviewers.

\bibliographystyle{ACM-Reference-Format}
\bibliography{refs}
\newpage
\pagenumbering{roman}
\setcounter{page}{1}
\appendix

\section{Demographics}
\label{app:Demographics}

\begin{table*}[ht!]
  \caption{Gender distribution}
  \label{tab:dem gender}
  \begin{tabular}{lccccc}
    \toprule
    Gender & 2018 ACS & 2020 seg 1 &    2020 seg 2 &    2020 seg 3 &    2021 \\
    \midrule
    Female &    0.496 & 0.622 &     0.612 &     0.625 &     0.530 \\
    Male   &    0.504 & 0.345 &     0.337 &     0.340 &     0.432 \\
    Other  &          & 0.033 &     0.051 &     0.035 &     0.038 \\
    \bottomrule
  \end{tabular}
\end{table*}

\begin{table*}[ht!]
  \caption{District distribution}
  \label{tab:dem district}
  \begin{tabular}{lccccc}
    \toprule
    District & 2010 Census    &     2020 seg 1 &    2020 seg 2 &    2020 seg 3 &    2021 \\
    \midrule
    District  1 &   0.097 & 0.143 & 0.159 & 0.140 &  0.102 \\
    District  2 &   0.1 &   0.077 & 0.051 & 0.048 &  0.069 \\
    District  3 &   0.1 &   0.104 & 0.099 & 0.093 &  0.081 \\
    District  4 &   0.099 & 0.064 & 0.065 & 0.063 &  0.082 \\
    District  5 &   0.102 & 0.075 & 0.082 & 0.105 &  0.140 \\
    District  6 &   0.103 & 0.052 & 0.043 & 0.051 &  0.053 \\
    District  7 &   0.101 & 0.103 & 0.087 & 0.103 &  0.112 \\
    District  8 &   0.097 & 0.101 & 0.046 & 0.065 &  0.095 \\
    District  9 &   0.099 & 0.085 & 0.149 & 0.135 &  0.136 \\
    District 10 &   0.101 & 0.107 & 0.070 & 0.088 &  0.118 \\
    Other       &   - &     0.090 & 0.150 & 0.108 &  0.012 \\
    \bottomrule
  \end{tabular}
\end{table*}

\begin{table*}[ht!]
  \caption{Age distribution}
  \label{tab:dem age}
  \begin{tabular}{lccccc}
    \toprule
    Age Group & 2018 ACS &     2020 seg 1 &    2020 seg 2 &    2020 seg 3 &    2021 \\
    \midrule
    18-   &     0.197 & -   &   0.016 & 0.014 &        \\
    18-24 &     0.105 & 0.034 & 0.304 & 0.167 &  0.052 \\
    25-34 &     0.227 & 0.283 & 0.457 & 0.431 &  0.292 \\
    35-44 &     0.156 & 0.246 & 0.155 & 0.223 &  0.266 \\
    45-54 &     0.119 & 0.198 & 0.044 & 0.087 &  0.177 \\
    55-64 &     0.103 & 0.14 &  0.017 & 0.048 &  0.106 \\
    65+   &     0.094 & 0.098 & 0.008 & 0.030 &  0.107 \\
    \bottomrule
  \end{tabular}
\end{table*}

\begin{table*}[ht!]
  \caption{Race/Ethnicity distribution}
  \label{tab:dem race}
  \begin{tabular}{lccccc}
    \toprule
    Race/Ethnicity          & 2018 ACS &  2020 seg 1 &    2020 seg 2 &    2020 seg 3 &    2021 \\
    \midrule
    Asian alone                  & 0.076 & 0.030 & 0.074 & 0.068 &  0.050 \\
    Black/African American alone & 0.081 & 0.022 & 0.028 & 0.033 &  0.045 \\
    Latinx/hispanic              & 0.327 & 0.181 & 0.202 & 0.183 &  0.104 \\
    Other/multiple races         & 0.038 & 0.029 & 0.040 & 0.036 &  0.011 \\
    White alone                  & 0.488 & 0.738 & 0.656 & 0.680 &  0.790 \\
    \bottomrule
  \end{tabular}
\end{table*}

\begin{table*}[ht!]
  \caption{Home Ownership distribution}
  \label{tab:dem home}
  \begin{tabular}{lccccc}
    \toprule
    Home Ownership & 2018 ACS & 2020 seg 1 &    2020 seg 2 &    2020 seg 3 &    2021 \\
    \midrule
    Own     &   0.479 & 0.64 &  0.341 & 0.48 &  0.682 \\
    Rent    &   0.499 & 0.36 &  0.659 & 0.52 &  0.318 \\    
    \bottomrule
  \end{tabular}
\end{table*}

\begin{table*}[ht!]
  \caption{Income distribution by time segment}
  \label{tab:dem income}
  \begin{tabular}{lccccc}
    \toprule
    Income Group & 2018 ACS & 2020 seg 1 & 2020 seg 2 & 2020 seg 3 & 2021 \\
    \midrule
    \$ 35k or less  & 0.226 & 0.118 & 0.235 & 0.148 &  0.091 \\
    \$ 35-100k      & 0.408 & 0.462 & 0.451 & 0.444 &  0.434 \\    
    \$ 100-150k     & 0.170 & 0.227 & 0.168 & 0.202 &  0.218 \\
    \$ 150k or more & 0.197 & 0.193 & 0.147 & 0.206 &  0.257 \\
    \bottomrule
  \end{tabular}
\end{table*}

\newpage
\section{Outcomes 2020}
\label{app:outcomes 2020}

\begin{table*}[ht!]
  \caption{ Preferred Fee Change per Category, 2020}
  \label{tab:revenue 2020}
  \begin{tabular}{lccc}
    \toprule
    Fee Category &  No change &  Moderate increase &  Significant increase \\
    \midrule
    Animal Services Fees              & 0.508 & 0.411 & 0.081 \\
    Aquatic Fees                      & 0.633 & 0.304 & 0.063 \\
    EMS Transport Fees                & 0.875 & 0.080 & 0.045 \\
    Facility Rental Fees              & 0.546 & 0.372 & 0.083 \\
    Fire Permit \& Inspection Fees    & 0.385 & 0.434 & 0.180 \\
    Golf fees                         & 0.242 & 0.256 & 0.503 \\
    Parks and Recreation Program Fees & 0.655 & 0.292 & 0.054 \\
    Planning and Zoning Fees          & 0.465 & 0.342 & 0.193 \\
    Public Health Permit Fees         & 0.586 & 0.349 & 0.064 \\
    \bottomrule
  \end{tabular}
\end{table*}

\begin{table*}[ht!]
  \caption{Preferred Budget Change per Service Area, 2020}
  \label{tab:expenditure 2020}
  \begin{tabular}{lccccc}
    \toprule
    Service Area &  Significant reduction &  Moderate reduction &  No change &  Mod. increase &  Sign. increase \\
    \midrule
    Animal Services             & 0.035 & 0.033 & 0.391 & 0.136 & 0.405 \\
    Austin Fire Department      & 0.017 & 0.073 & 0.475 & 0.426 & 0.010 \\
    Austin Police Department    & 0.598 & 0.317 & 0.049 & 0.034 & 0.001 \\
    Austin Public Health        & 0.005 & 0.012 & 0.098 & 0.313 & 0.573 \\
    Austin Public Library       & 0.013 & 0.035 & 0.272 & 0.350 & 0.330 \\
    Emergency Medical Services  & 0.004 & 0.014 & 0.255 & 0.461 & 0.265 \\
    Municipal Court             & 0.060 & 0.066 & 0.514 & 0.166 & 0.194 \\
    NHCD                        & 0.023 & 0.016 & 0.146 & 0.069 & 0.745 \\
    Other                       & 0.036 & 0.078 & 0.554 & 0.226 & 0.106 \\
    Parks and Recreation        & 0.009 & 0.035 & 0.308 & 0.498 & 0.151 \\
    Planning and Zoning         & 0.074 & 0.069 & 0.562 & 0.079 & 0.216 \\
    \bottomrule
  \end{tabular}
\end{table*}

\newpage
\begin{table*}[ht!]
  \caption{ Aggregated City Budget from responses, 2020}
  \label{tab:expenditure 2020 knapsack}
  \begin{tabular}{lrrr}
    \toprule
    Service Area & Original Budget & Proposed Change & Change\% \\
    \midrule
      Austin Police Department &  434,475,745.00 &  -13,000,000.00 &  -2.99\% \\
        Austin Fire Department &  200,701,475.00 &     +250,000.00 &  +0.12\% \\
          Parks and Recreation &   98,394,261.00 &   +1,000,000.00 &  +1.02\% \\
    Emergency Medical Services &   93,068,228.00 &   +2,000,000.00 &  +2.15\% \\
          Austin Public Health &   85,926,146.00 &   +4,750,000.00 &  +5.53\% \\
         Austin Public Library &   54,685,661.00 &   +1,250,000.00 &  +2.29\% \\
                         Other &   49,699,345.00 &           - &  - \\
               Municipal Court &   31,510,968.00 &           - &  - \\
               Animal Services &   15,552,062.00 &     +500,000.00 &  +3.22\% \\
                          NHCD &   14,829,857.00 &   +3,250,000.00 & +21.92\% \\
           Planning and Zoning &    9,732,705.00 &           - &  - \\
       \bottomrule
  \end{tabular}
\end{table*}

\clearpage
\section{Outcomes 2021}
\label{app:outcomes 2021}
\begin{table*}[ht!]
  \caption{Preferred Fee Change per Service Area, 2021}
  \label{tab:revenue 2021}
    \begin{tabular}{lccc}
    \toprule
    {Fee Category} &  No change &  Moderate increase &  Significant increase \\
    \midrule
    Animal Services Fees              & 0.556 & 0.353 & 0.091 \\
    Aquatic Fees                      & 0.631 & 0.305 & 0.064 \\
    EMS Transport Fees                & 0.794 & 0.160 & 0.046 \\
    Facility Rental Fees              & 0.462 & 0.407 & 0.131 \\
    Fire Permit \& Inspection Fees    & 0.407 & 0.423 & 0.170 \\
    Golf fees                         & 0.351 & 0.311 & 0.338 \\
    Parks and Recreation Program Fees & 0.569 & 0.358 & 0.072 \\
    Planning and Zoning Fees          & 0.537 & 0.280 & 0.183 \\
    Public Health Permit Fees         & 0.559 & 0.359 & 0.082 \\
    \bottomrule
    \end{tabular}
\end{table*}

\begin{table*}[ht!]
  \caption{Preferred Budget Change per Service Area, 2021}
  \label{tab:expenditure 2021}
  \begin{tabular}{lccccc}
    \toprule
    {Service Area} &  Significant decrease &  Moderate decrease &  No change &  Moderate increase &  Significant increase \\
    \midrule
    Animal Services            & 0.074 & 0.102 & 0.499 & 0.261 & 0.065 \\
    Austin Fire Department     & 0.040 & 0.065 & 0.481 & 0.343 & 0.071 \\
    Austin Police Department   & 0.326 & 0.130 & 0.139 & 0.125 & 0.281 \\
    Austin Public Health       & 0.078 & 0.081 & 0.268 & 0.322 & 0.251 \\
    Austin Public Library      & 0.095 & 0.115 & 0.427 & 0.258 & 0.105 \\
    Emergency Medical Services & 0.016 & 0.034 & 0.411 & 0.374 & 0.165 \\
    Housing and Planning       & 0.142 & 0.094 & 0.255 & 0.257 & 0.252 \\
    Municipal Court            & 0.078 & 0.121 & 0.610 & 0.162 & 0.028 \\
    Other                      & 0.153 & 0.144 & 0.532 & 0.140 & 0.032 \\
    Parks and Recreation       & 0.028 & 0.075 & 0.416 & 0.339 & 0.143 \\
    Public Safety Support      & 0.225 & 0.084 & 0.267 & 0.212 & 0.212 \\
    \bottomrule
\end{tabular}
\end{table*}

\newpage
\section{Outcomes Over Time}
\label{app:outcomes time}

\begin{table*}[ht!]
  \caption{Support for increasing property tax by time segment}
  \label{tab:tax multi}
  \begin{tabular}{lccccc}
    \toprule
    Income Group & 2020 & 2020 seg 1 & 2020 seg 2 & 2020 seg 3 & 2021 \\
    \midrule
    No         &  0.354 & 0.528 & 0.336 & 0.434 &  0.654 \\
    Yes        &  0.501 & 0.410 & 0.515 & 0.432 &  0.278 \\    
    No Opinion &  0.145 & 0.062 & 0.149 & 0.134 &  0.068 \\
    \bottomrule
  \end{tabular}
\end{table*}

\begin{table*}[ht!]
  \caption{Support for any Increase of Fees by time segment}
  \label{tab:revenue multi}
  \begin{tabular}{lccc}
    \toprule
    Revenue Category & 2020 seg 1 & 2020 seg 2 & 2020 seg 3 \\
    \midrule
    Animal Service Fees                 &  0.553 &  0.452 &  0.483 \\
    Aquatic Fees                        &  0.499 &  0.342 &  0.385 \\
    EMS Transport Fees                  &  0.213 &  0.114 &  0.135 \\
    Fire Permit \& Inspection Fees      &  0.728 &  0.549 &  0.604 \\
    Golf Fees                           &  0.792 &  0.712 &  0.730 \\
    Public Health Permit Fees    &  0.573 &  0.366 &  0.427 \\
    Parks and Recreation Program Fees   &  0.533 &  0.317 &  0.361 \\
    Facility Rental Fees                &  0.674 &  0.404 &  0.478 \\
    Planning and Zoning Fees            &  0.631 &  0.458 &  0.499 \\
    \bottomrule  \end{tabular}
\end{table*}

\begin{table*}[ht!]
  \caption{Support for any Increase of Budget by time segment}
  \label{tab:expenditure multi}
  \begin{tabular}{lccc}
    \toprule
    Department & 2020 seg 1 & 2020 seg 2 & 2020 seg 3 \\
    \midrule
    Animal Services             &  0.197 &  0.557 &  0.493 \\
    Austin Fire Department      &  0.260 &  0.446 &  0.404 \\
    Austin Police Department    &  0.229 &  0.027 &  0.059 \\
    Austin Public Health        &  0.537 &  0.899 &  0.850 \\
    Austin Public Library       &  0.284 &  0.702 &  0.605 \\
    Emergency Medical Services  &  0.399 &  0.743 &  0.674 \\
    Municipal Court             &  0.097 &  0.370 &  0.332 \\
    NHCD                        &  0.392 &  0.828 &  0.792 \\
    Other                       &  0.120 &  0.337 &  0.327 \\
    Parks and Recreation        &  0.366 &  0.661 &  0.613 \\
    Planning and Zoning         &  0.104 &  0.301 &  0.281 \\
    \bottomrule  \end{tabular}
\end{table*}

\begin{table*}[ht!]
  \caption{Support for any Decrease of Budget}
  \label{tab:expenditure decrease multi}
    \begin{tabular}{lrrrrl}
    \toprule
    Department &      2020 seg 1 &      2020 seg 2 &      2020 seg 3  \\
    \midrule
    Animal Services             &  0.227 &  0.061 &  0.083     \\
    Austin Fire Department      &  0.136 &  0.087 &  0.093       \\
    Austin Police Department    &  0.427 &  0.934 &  0.872      \\
    Austin Public Health        &  0.085 &  0.013 &  0.024       \\
    Austin Public Library       &  0.216 &  0.040 &  0.070       \\
    Emergency Medical Services  &  0.054 &  0.017 &  0.022      \\
    Municipal Court             &  0.263 &  0.123 &  0.129      \\
    NHCD                        &  0.214 &  0.032 &  0.058      \\
    Other                       &  0.370 &  0.106 &  0.125      \\
    Parks and Recreation        &  0.146 &  0.040 &  0.055     \\
    Planning and Zoning         &  0.326 &  0.139 &  0.143     \\
    \bottomrule
    \end{tabular}
\end{table*}

\clearpage
\subsection{Age Adjusted results for 2020 over time}
\label{app:outcomes 2020 time age}
\begin{table*}[ht!]
  \caption{Support for Increase of Fees, Age Adjusted (2020 seg  1), n=642}
  \label{tab:revenue 2020-1 age}
  \begin{tabular}{lccc}
    \toprule
    {} &  No change &  Moderate increase &  Significant increase \\
    \midrule
    Animal Services Fees              &      0.416 & 0.456 & 0.128 \\
    Aquatic Fees                      &      0.501 & 0.377 & 0.123 \\
    EMS Transport Fees                &      0.793 & 0.147 & 0.060 \\
    Facility Rental Fees              &      0.352 & 0.487 & 0.160 \\
    Fire Permit \& Inspection Fees     &      0.265 & 0.440 & 0.295 \\
    Golf fees                         &      0.202 & 0.315 & 0.483 \\
    Parks and Recreation Program Fees &      0.467 & 0.422 & 0.111 \\
    Planning and Zoning Fees          &      0.317 & 0.337 & 0.345 \\
    Public Health Permit Fees         &      0.426 & 0.461 & 0.113 \\
    \bottomrule
    \end{tabular}
\end{table*}

\begin{table*}[ht!]
  \caption{Support for Increase of Fees, Age Adjusted (2020 seg 2), n=29,544}
  \label{tab:revenue 2020-2 age}
  \begin{tabular}{lccc}
    \toprule
    {} &  No change &  Moderate increase &  Significant increase \\
    \midrule
    Animal Services Fees              &      0.521 & 0.405 & 0.075 \\
    Aquatic Fees                      &      0.637 & 0.302 & 0.061 \\
    EMS Transport Fees                &      0.857 & 0.099 & 0.044 \\
    Facility Rental Fees              &      0.528 & 0.386 & 0.087 \\
    Fire Permit \& Inspection Fees     &      0.379 & 0.430 & 0.191 \\
    Golf fees                         &      0.246 & 0.269 & 0.484 \\
    Parks and Recreation Program Fees &      0.637 & 0.310 & 0.053 \\
    Planning and Zoning Fees          &      0.457 & 0.351 & 0.192 \\
    Public Health Permit Fees         &      0.580 & 0.357 & 0.063 \\
    \bottomrule
    \end{tabular}
\end{table*}

\begin{table*}[ht!]
  \caption{Support for Increase of Fees, Age Adjusted (2020 seg 3), n=4964}
  \label{tab:revenue 2020-3 age}
  \begin{tabular}{lccc}
    \toprule
    {} &  No change &  Moderate increase &  Significant increase \\
    \midrule
    Animal Services Fees              & 0.496 & 0.419 & 0.085 \\
    Aquatic Fees                      & 0.592 & 0.341 & 0.066 \\
    EMS Transport Fees                & 0.846 & 0.116 & 0.038 \\
    Facility Rental Fees              & 0.465 & 0.440 & 0.095 \\
    Fire Permit \& Inspection Fees    & 0.337 & 0.466 & 0.197 \\
    Golf fees                         & 0.225 & 0.318 & 0.457 \\
    Parks and Recreation Program Fees & 0.604 & 0.332 & 0.064 \\
    Planning and Zoning Fees          & 0.430 & 0.381 & 0.190 \\
    Public Health Permit Fees         & 0.527 & 0.400 & 0.073 \\
    \bottomrule
    \end{tabular}
\end{table*}

\begin{table*}[ht!]
  \caption{Support for Change of Budget, Age Adjusted (2020 seg 1), n=642}
  \label{tab:expenditure 2020-1 age}
  \begin{tabular}{lccccc}
    \toprule
    {} &  Significant reduction &  Moderate reduction &  No change &  Mod. increase &  Sign. increase \\
    \midrule
    Animal Services            & 0.064 & 0.131 & 0.557 & 0.135 & 0.112 \\
    Austin Fire Department     & 0.007 & 0.160 & 0.575 & 0.257 & 0.001 \\
    Austin Police Department   & 0.132 & 0.359 & 0.332 & 0.175 & 0.001 \\
    Austin Public Health       & 0.010 & 0.046 & 0.328 & 0.464 & 0.152 \\
    Austin Public Library      & 0.031 & 0.143 & 0.453 & 0.291 & 0.082 \\
    Emergency Medical Services & 0.001 & 0.051 & 0.483 & 0.412 & 0.053 \\
    Municipal Court            & 0.038 & 0.227 & 0.624 & 0.071 & 0.039 \\
    NHCD                       & 0.089 & 0.081 & 0.383 & 0.135 & 0.313 \\
    Other                      & 0.057 & 0.305 & 0.488 & 0.125 & 0.025 \\
    Parks and Recreation       & 0.011 & 0.144 & 0.447 & 0.349 & 0.049 \\
    Planning and Zoning        & 0.126 & 0.169 & 0.563 & 0.043 & 0.099 \\
    \bottomrule
    \end{tabular}
\end{table*}

\begin{table*}[ht!]
  \caption{Support for Change of Budget, Age Adjusted (2020 seg 2), n=29545}
  \label{tab:expenditure 2020-2 age}
  \begin{tabular}{lccccc}
    \toprule
    {} &  Significant reduction &  Moderate reduction &  No change &  Mod. increase &  Sign. increase \\
    \midrule
    Animal Services            & 0.042 & 0.046 & 0.405 & 0.140 & 0.368 \\
    Austin Fire Department     & 0.015 & 0.072 & 0.479 & 0.425 & 0.010 \\
    Austin Police Department   & 0.536 & 0.336 & 0.071 & 0.056 & 0.001 \\
    Austin Public Health       & 0.005 & 0.021 & 0.126 & 0.324 & 0.524 \\
    Austin Public Library      & 0.015 & 0.052 & 0.291 & 0.339 & 0.304 \\
    Emergency Medical Services & 0.003 & 0.017 & 0.292 & 0.454 & 0.233 \\
    Municipal Court            & 0.059 & 0.079 & 0.530 & 0.160 & 0.172 \\
    NHCD                       & 0.030 & 0.022 & 0.172 & 0.086 & 0.689 \\
    Other                      & 0.043 & 0.101 & 0.550 & 0.209 & 0.097 \\
    Parks and Recreation       & 0.009 & 0.043 & 0.325 & 0.482 & 0.142 \\
    Planning and Zoning        & 0.087 & 0.087 & 0.559 & 0.075 & 0.191 \\
    \bottomrule
    \end{tabular}
\end{table*}

\begin{table*}[ht!]
  \caption{Support for Change of Budget, Age Adjusted (2020 seg 3), n=4965}
  \label{tab:expenditure 2020-3 age}
  \begin{tabular}{lccccc}
    \toprule
    {} &  Significant reduction &  Moderate reduction &  No change &  Mod. increase &  Sign. increase \\
    \midrule
    Animal Services            & 0.044 & 0.042 & 0.436 & 0.130 & 0.347 \\
    Austin Fire Department     & 0.021 & 0.076 & 0.505 & 0.389 & 0.010 \\
    Austin Police Department   & 0.557 & 0.286 & 0.085 & 0.071 & 0.001 \\
    Austin Public Health       & 0.005 & 0.020 & 0.146 & 0.291 & 0.537 \\
    Austin Public Library      & 0.021 & 0.062 & 0.341 & 0.325 & 0.252 \\
    Emergency Medical Services & 0.008 & 0.018 & 0.317 & 0.422 & 0.234 \\
    Municipal Court            & 0.058 & 0.075 & 0.545 & 0.151 & 0.172 \\
    NHCD                       & 0.034 & 0.034 & 0.168 & 0.072 & 0.691 \\
    Other                      & 0.044 & 0.102 & 0.547 & 0.197 & 0.110 \\
    Parks and Recreation       & 0.016 & 0.050 & 0.354 & 0.444 & 0.136 \\
    Planning and Zoning        & 0.078 & 0.087 & 0.565 & 0.069 & 0.201 \\
    \bottomrule
    \end{tabular}
\end{table*}

\clearpage
\newpage
\section{Follow-up Survey 2021}
\label{app:followup 2021}
\begin{figure}[ht!]
    \centering
    \begin{minipage}{0.48 \linewidth}
        \centering
        \includegraphics[width=0.95\linewidth, alt={Table with 3 revenue scenarios as presented to participants in the 2021 follow-up survey. Background colors indicate direction of change.}]{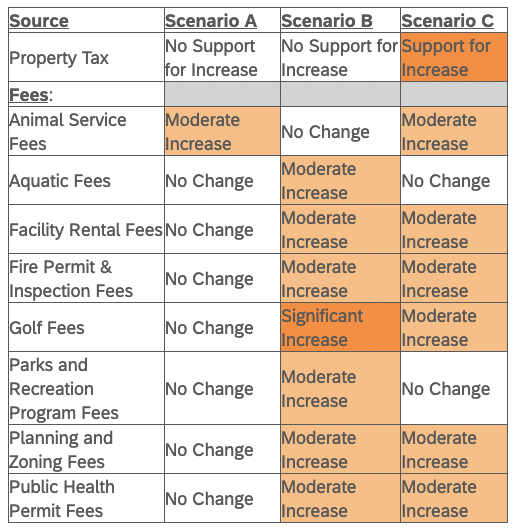}
        \caption{The 3 revenue scenarios as presented to participants in the 2021 follow-up survey. The order of the scenarios was randomized.}
        \label{fig:screenshot followup 2021 rev}
    \end{minipage}
    \begin{minipage}{0.48\linewidth}
        \centering
        \includegraphics[width=0.95\linewidth, alt={Table with 3 expenditure scenarios as presented to participants in the 2021 follow-up survey. Background colors indicate direction of change.}]{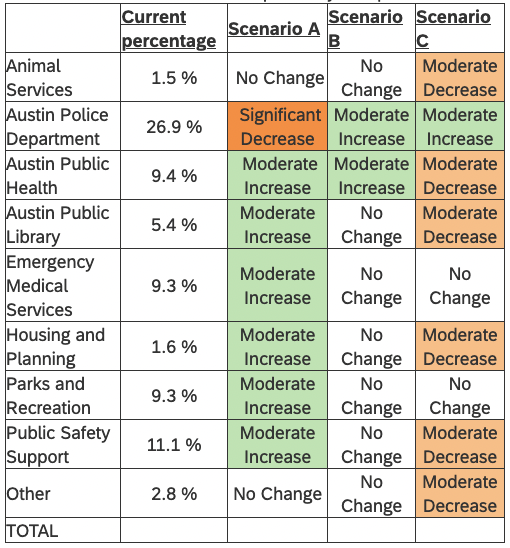}
        \caption{The 3 expenditure scenarios as presented to participants in the 2021 follow-up survey. The order of the scenarios was randomized.}
        \label{fig:screenshot followup 2021 exp}
    \end{minipage}
\end{figure}

Three scenarios were designed based on aggregated clusters from the available 2021 responses. Fig.~\ref{fig:screenshot followup 2021 rev} and \ref{fig:screenshot followup 2021 exp} display them the way they were presented to the survey takers. 

\begin{enumerate}
    \item Scenario rev-A is no change in property tax, and no change in service fees, except for a moderate increase in animal service fees
    \item Scenario rev-B is no change in property tax, and a moderate increase in all service fees, except no increase for animal service fees and a significant increase for golf fees.
    \item Scenario rev-C is an increase in property tax, and a moderate increase in all service fees except no increase in aquatic fees and park and recreation program fees.
    \item Scenario exp-A is a significant decrease for the police department, and a moderate increase in all other departments, except no change to animal services and ‘other’.
    \item Scenario exp-B is a moderate increase for the police department, and Public Health, but no change to other departments.
    \item Scenario exp-C is a moderate increase for the police department, and a moderate decrease for all other departments except for Emergency Medical Services and Parks and Recreation. 
\end{enumerate}

\begin{table*}[ht!]
  \caption{Ranked preference per revenue (n=127) and expenditure scenario (n=135)}
  \label{tab:followup 2021 rankings}
    \begin{tabular}{lrrrrrr}
    \toprule
    {Rank} &  rev-A &  rev-B &  rev-C &  exp-A &  exp-B &  exp-C \\
    \midrule
    1 &          36 &          60 &          31 &          80 &          37 &          18 \\
    2 &          40 &          52 &          35 &          14 &          87 &          34 \\
    3 &          51 &          15 &          61 &          41 &          11 &          83 \\
    \bottomrule
    \end{tabular}
\end{table*}
\newpage
\section{Clustering}

\begin{figure}[ht!]
    \centering
    \begin{minipage}{0.48 \linewidth}
        \centering
        \includegraphics[width=0.95\linewidth]{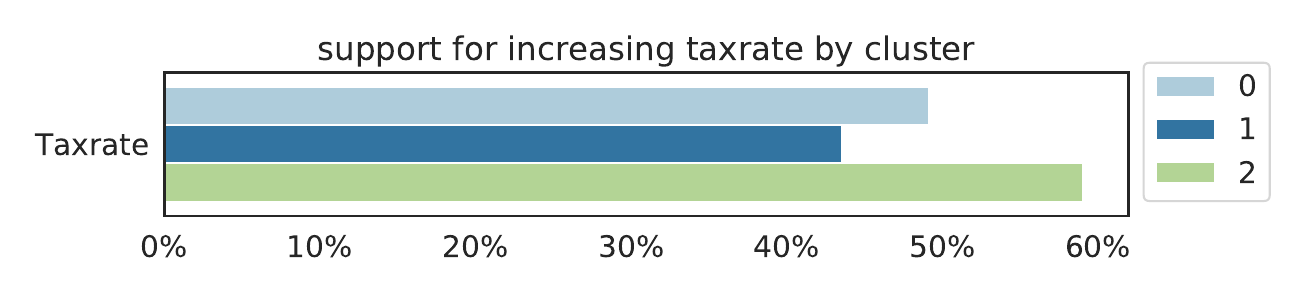}
        \caption{Support for increase in property tax, by cluster (2020)}
        \label{fig:tax 2020 3clusters}
    \end{minipage}
    \begin{minipage}{0.48\linewidth}
        \centering
        \includegraphics[width=0.95\linewidth]{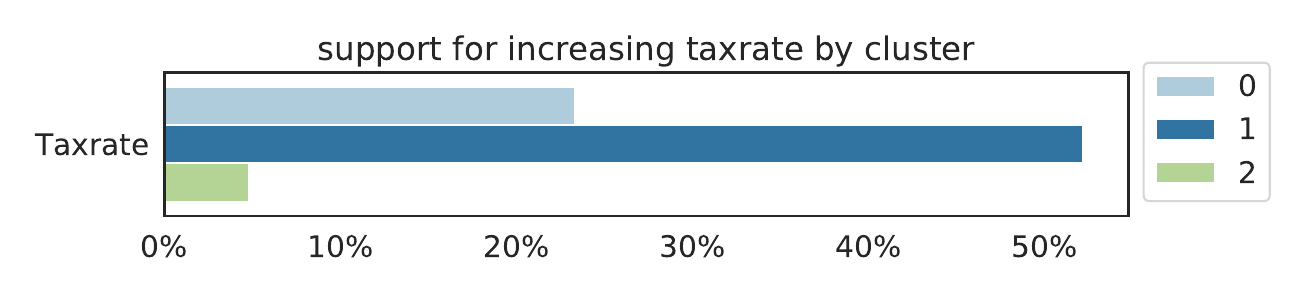}
        \caption{Support for increase in property tax, by cluster (2021)}
        \label{fig:tax 2021 3clusters}
    \end{minipage}
\end{figure}

\label{app:clustering}
\begin{figure}[ht!]
    \centering
    \begin{minipage}{0.48 \linewidth}
        \centering
        \label{fig:bootstrap 2020 2clusters}
        \includegraphics[width=0.95\linewidth, alt={Horizontal bar chart displaying the distribution of the cluster means per expenditure/revenue area after bootstrapping}]{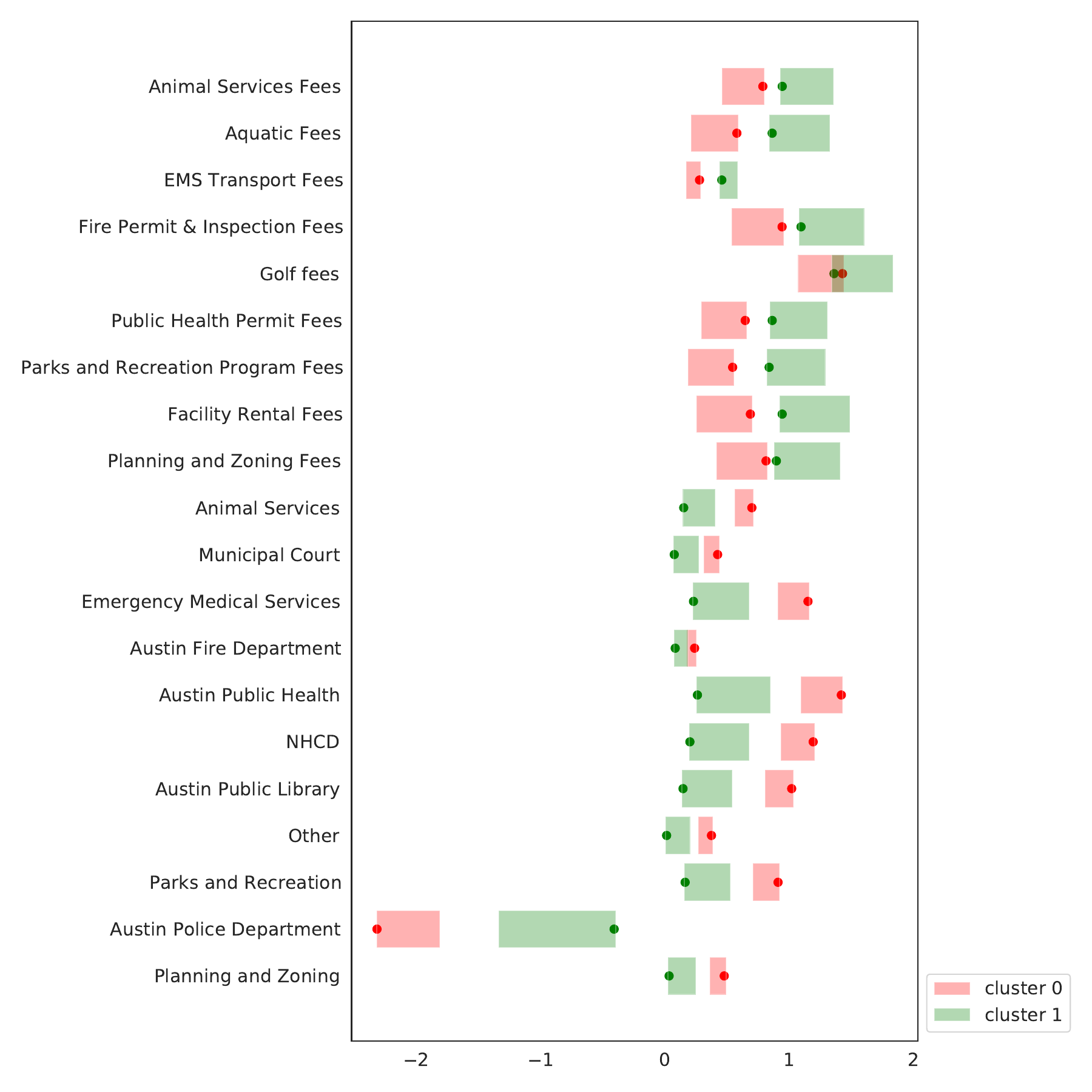}
        \caption{2020 centroid budget shift per service area, for 2 clusters with 0.95 confidence interval}
    \end{minipage}
    \begin{minipage}{0.48\linewidth}
        \centering
        \label{fig:bootstrap 2020 4clusters}
        \includegraphics[width=0.95\linewidth, alt={Horizontal bar chart displaying the distribution of the cluster means per expenditure/revenue area after bootstrapping}]{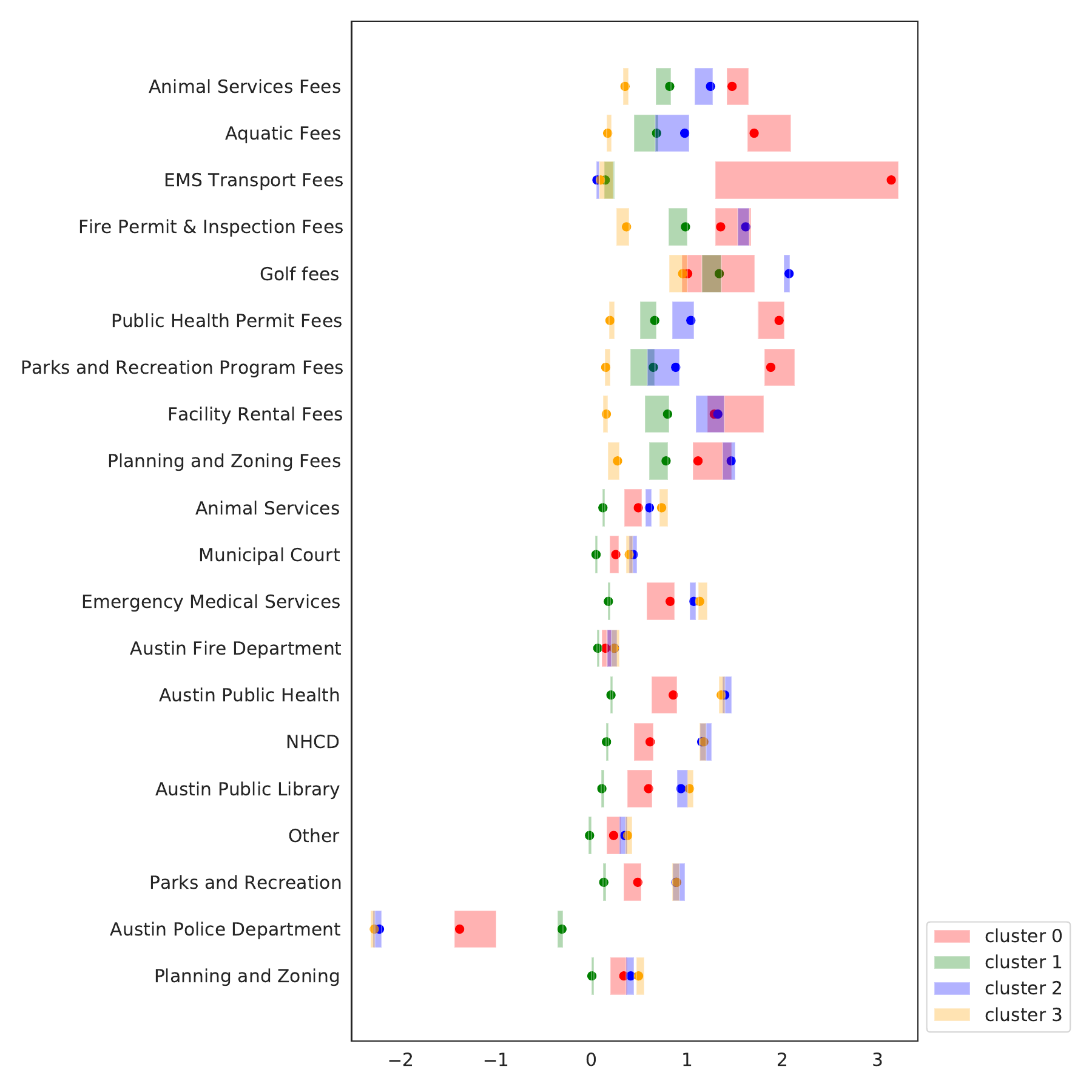}
        \caption{2020 centroid budget shift per service area, for 4 clusters with 0.95 confidence interval}
    \end{minipage}
\end{figure}

\begin{figure}[ht!]
    \centering
    \begin{minipage}{0.48 \linewidth}
        \centering
        \includegraphics[width=0.95\linewidth, alt={Horizontal bar chart displaying the distribution of the cluster means per expenditure/revenue area after bootstrapping}]{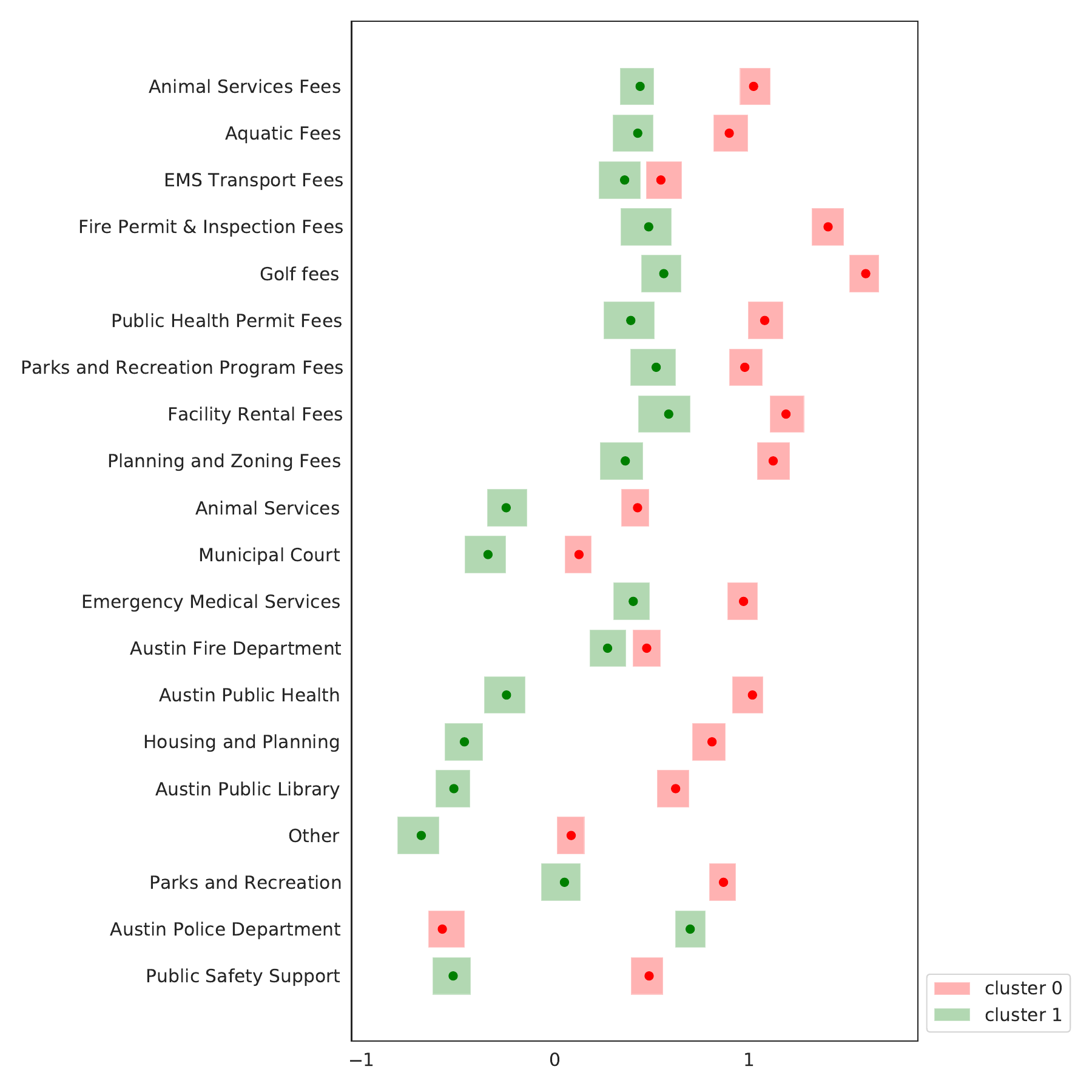}
        \caption{2021 centroid budget shift per service area, for 2 clusters with 0.95 confidence interval}
        \label{fig:bootstrap 2021 2clusters}
    \end{minipage}
    \begin{minipage}{0.48\linewidth}
        \centering
        \includegraphics[width=0.95\linewidth, alt={Horizontal bar chart displaying the distribution of the cluster means per expenditure/revenue area after bootstrapping}]{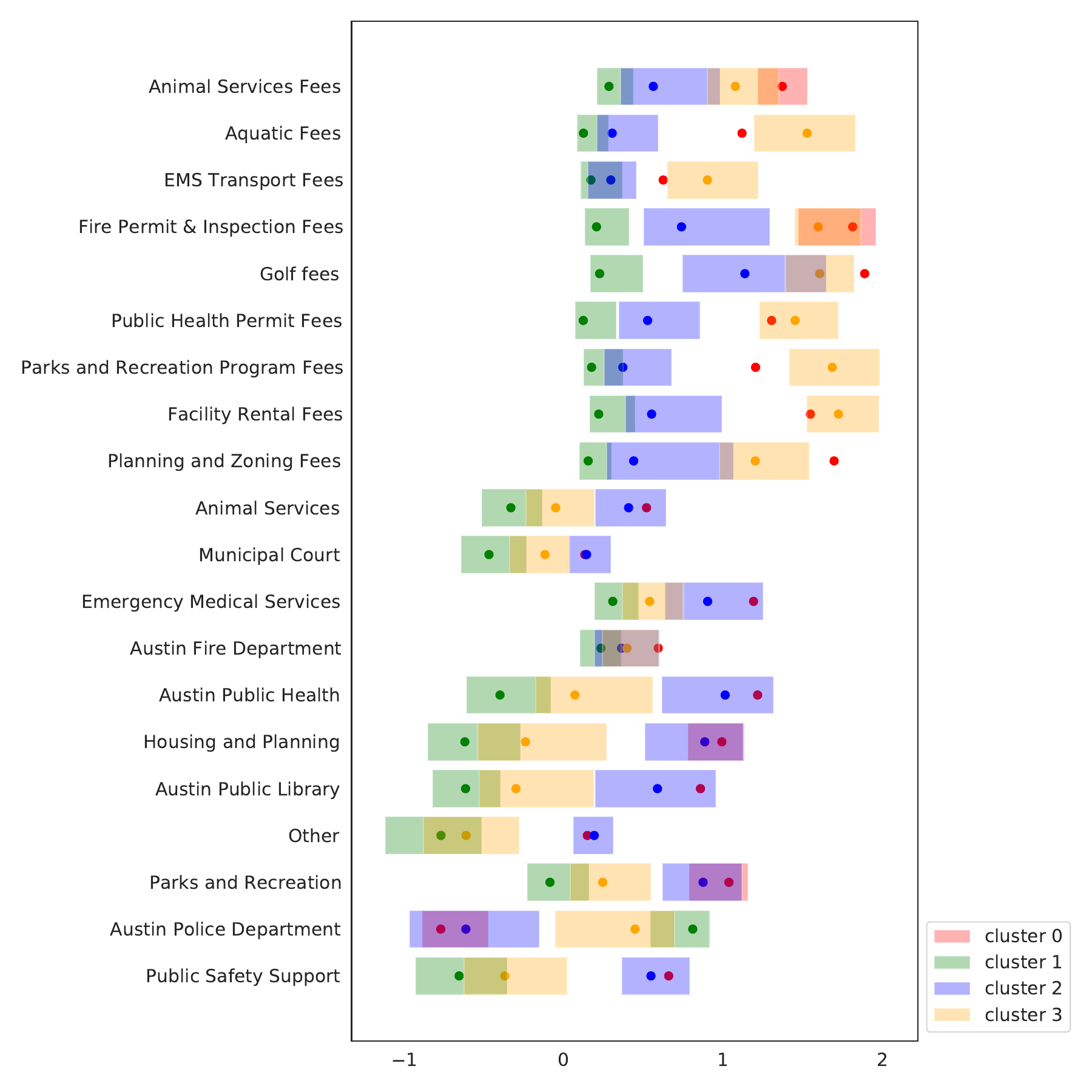}
        \caption{2021 centroid budget shift per service area, for 4 clusters with 0.95 confidence interval}
        \label{fig:bootstrap 2021 4clusters}
    \end{minipage}
\end{figure}

\section{PB Clustering}
\label{app:sec:pb clustering}

\begin{longtable}[ht]{lrr}
\toprule
{} &  votes &  clusters \\
election ID &        &           \\
\midrule
4           &    121 &         5 \\
5           &    117 &         4 \\
14          &    130 &         2 \\
17          &    769 &         4 \\
79          &    361 &         2 \\
93          &    724 &         4 \\
103         &    186 &         3 \\
111         &    388 &         2 \\
112         &   1093 &         5 \\
113         &   1021 &         5 \\
114         &    423 &         4 \\
116         &    190 &         3 \\
119         &    141 &         7 \\
121         &    119 &         3 \\
122         &    177 &         1 \\
126         &    265 &         4 \\
128         &    224 &         2 \\
130         &    495 &         7 \\
151         &     88 &         1 \\
170         &    254 &         5 \\
171         &    244 &         2 \\
172         &    389 &         3 \\
173         &    246 &         4 \\
174         &    430 &         3 \\
181         &     83 &         2 \\
182         &    119 &         4 \\
187         &    488 &         4 \\
192         &    169 &         2 \\
194         &    190 &         7 \\
196         &    107 &         2 \\
206         &    262 &         2 \\
218         &    191 &         3 \\
228         &    119 &         1 \\
229         &    207 &         3 \\
230         &    120 &         3 \\
232         &    102 &         4 \\
233         &    150 &         7 \\
234         &    196 &         4 \\
235         &    291 &         4 \\
236         &    104 &         3 \\
241         &    201 &         1 \\
246         &    256 &         4 \\
248         &    152 &         4 \\
249         &    470 &         4 \\
250         &    516 &         7 \\
251         &    168 &         5 \\
255         &    457 &         2 \\
256         &    945 &         1 \\
\bottomrule
    \caption{Optimal number of clusters for Participatory Budgeting elections with knapsack voting on the Stanford Participatory Budgeting Platform, using the gap statistic. Only cleaned data is used.}
    \label{app:tab:optimal pb clusters}
\end{longtable}

\newpage
\subsection{Progression of votes per cluster}
Progression of votes from each cluster for Participatory Bdugeting data. The colored lines show the fraction of votes in that cluster that have been cast so far. The dashed and dotted lines show respectively 1 and 2 standard deviations from the expected fraction. Only displaying plots for elections where the optimal number of clusters is at least 3, and the smallest cluster is at least 50 votes.

\begin{figure*}[ht]
    \centering
    \begin{minipage}{0.48 \linewidth}
        \centering
        \includegraphics[width=0.95\linewidth]{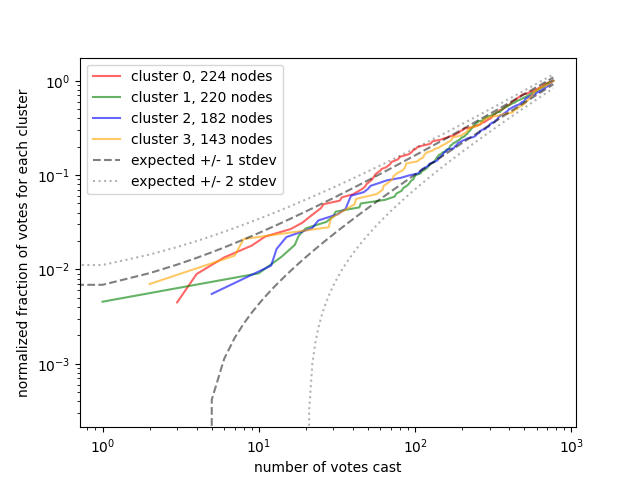}
        \setlength{\abovecaptionskip}{0pt}
        \caption{Progression of votes from each cluster for PB election 17.}
    \end{minipage}\hfill
    \begin{minipage}{0.48 \linewidth}
        \centering
        \includegraphics[width=0.95\linewidth]{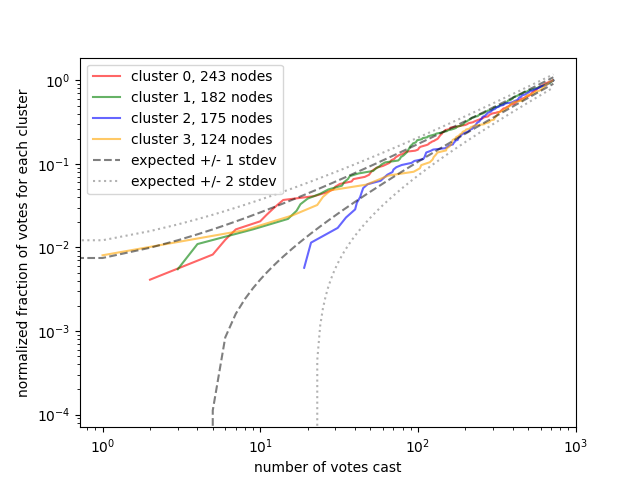}
        \setlength{\abovecaptionskip}{0pt}
        \caption{Progression of votes from each cluster for PB election 93.}
    \end{minipage}
\end{figure*}

\begin{figure*}[ht]
    \centering
    \begin{minipage}{0.48 \linewidth}
        \centering
        \includegraphics[width=0.95\linewidth]{figs/opinionChange/clusterplots/clusterfraction_over_time_PB_election_103.png}
        \setlength{\abovecaptionskip}{0pt}
        \caption{Progression of votes from each cluster for PB election 103.}
    \end{minipage}\hfill
    \begin{minipage}{0.48 \linewidth}
        \centering
        \includegraphics[width=0.95\linewidth]{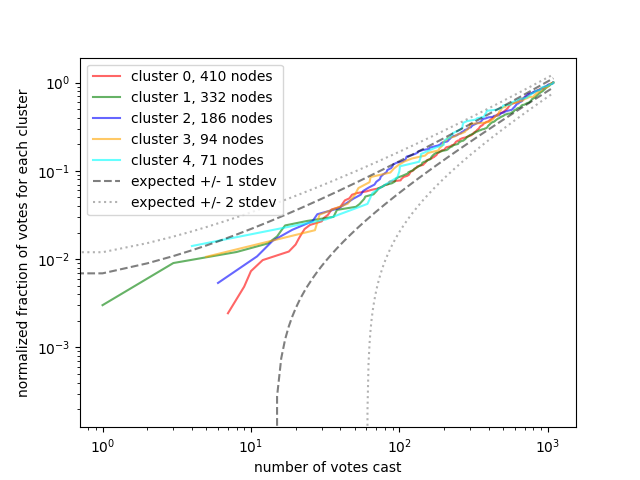}
        \setlength{\abovecaptionskip}{0pt}
        \caption{Progression of votes from each cluster for PB election 112.}
    \end{minipage}
\end{figure*}

\begin{figure*}[ht]
    \centering
    \begin{minipage}{0.48 \linewidth}
        \centering
        \includegraphics[width=0.95\linewidth]{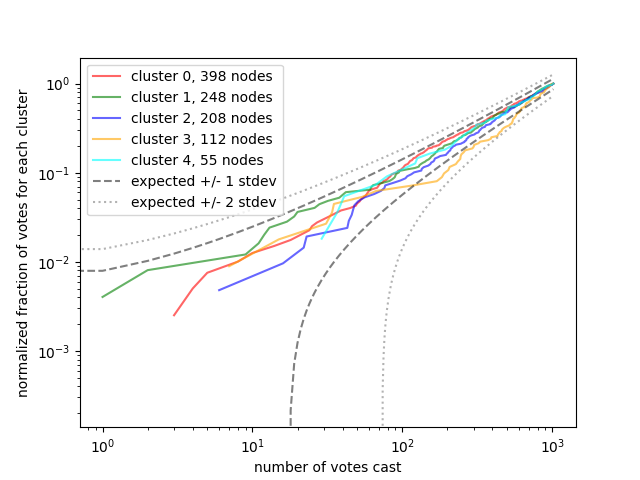}
        \setlength{\abovecaptionskip}{0pt}
        \caption{Progression of votes from each cluster for PB election 113.}
    \end{minipage}\hfill
    \begin{minipage}{0.48 \linewidth}
        \centering
        \includegraphics[width=0.95\linewidth]{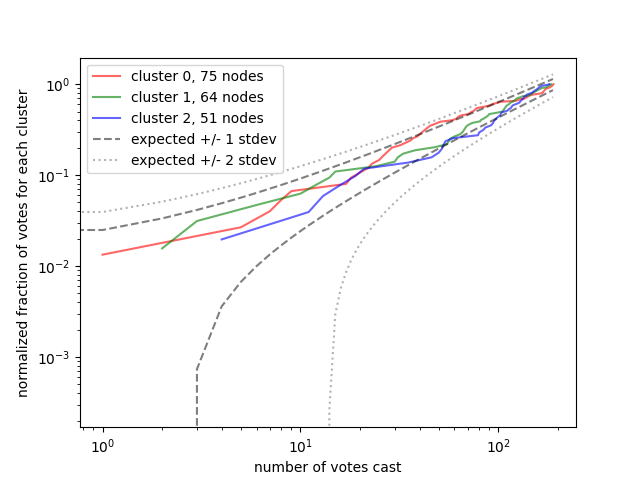}
        \setlength{\abovecaptionskip}{0pt}
        \caption{Progression of votes from each cluster for PB election 116.}
    \end{minipage}
\end{figure*}

\begin{figure*}[ht]
    \centering
    \begin{minipage}{0.48 \linewidth}
        \centering
        \includegraphics[width=0.95\linewidth]{figs/opinionChange/clusterplots/clusterfraction_over_time_PB_election_130.png}
        \setlength{\abovecaptionskip}{0pt}
        \caption{Progression of votes from each cluster for PB election 130.}
    \end{minipage}\hfill
    \begin{minipage}{0.48 \linewidth}
        \centering
        \includegraphics[width=0.95\linewidth]{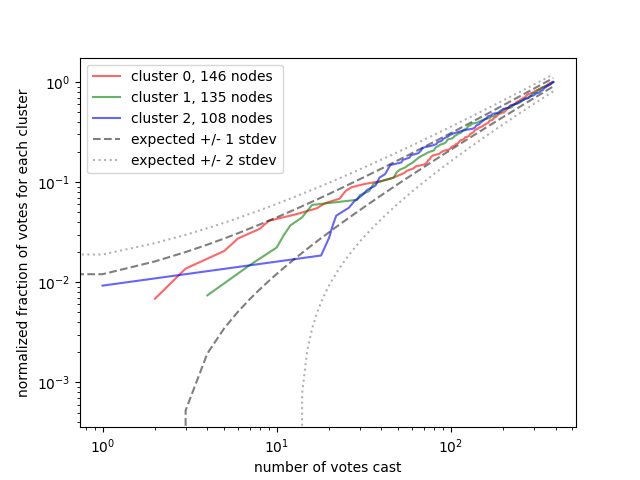}
        \setlength{\abovecaptionskip}{0pt}
        \caption{Progression of votes from each cluster for PB election 172.}
    \end{minipage}
\end{figure*}

\begin{figure*}[ht]
    \centering
    \begin{minipage}{0.48 \linewidth}
        \centering
        \includegraphics[width=0.95\linewidth]{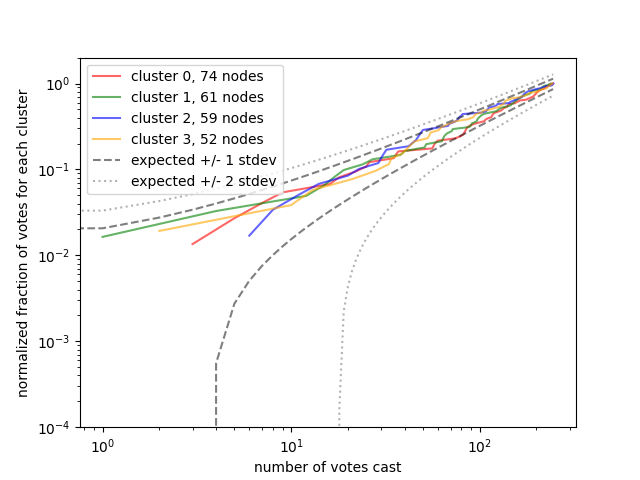}
        \setlength{\abovecaptionskip}{0pt}
        \caption{Progression of votes from each cluster for PB election 173.}
    \end{minipage}\hfill
    \begin{minipage}{0.48 \linewidth}
        \centering
        \includegraphics[width=0.95\linewidth]{figs/opinionChange/clusterplots/clusterfraction_over_time_PB_election_174.png}
        \setlength{\abovecaptionskip}{0pt}
        \caption{Progression of votes from each cluster for PB election 174.}
    \end{minipage}
\end{figure*}

\begin{figure*}[ht]
    \centering
    \begin{minipage}{0.48 \linewidth}
        \centering
        \includegraphics[width=0.95\linewidth]{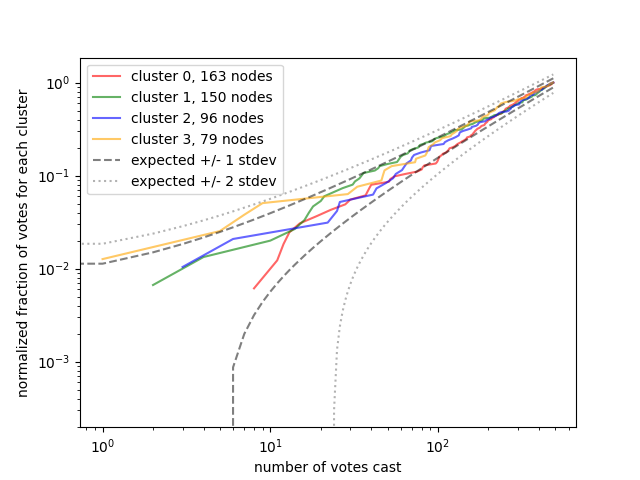}
        \setlength{\abovecaptionskip}{0pt}
        \caption{Progression of votes from each cluster for PB election 187.}
    \end{minipage}\hfill
    \begin{minipage}{0.48 \linewidth}
        \centering
        \includegraphics[width=0.95\linewidth]{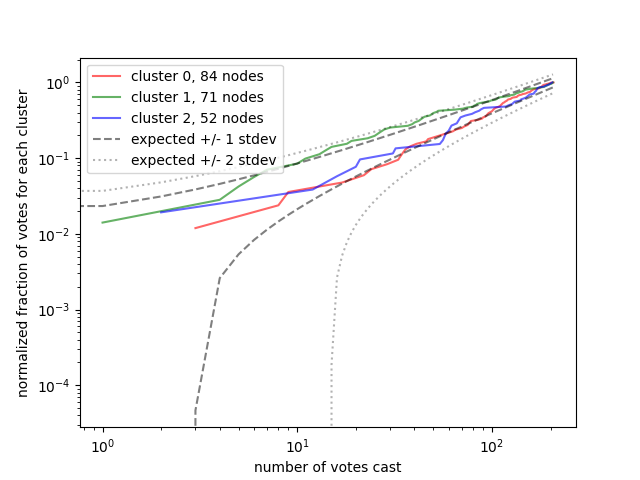}
        \setlength{\abovecaptionskip}{0pt}
        \caption{Progression of votes from each cluster for PB election 229.}
    \end{minipage}
\end{figure*}

\begin{figure*}[ht]
    \centering
    \begin{minipage}{0.48 \linewidth}
        \centering
        \includegraphics[width=0.95\linewidth]{figs/opinionChange/clusterplots/clusterfraction_over_time_PB_election_246.png}
        \setlength{\abovecaptionskip}{0pt}
        \caption{Progression of votes from each cluster for PB election 246.}
    \end{minipage}\hfill
    \begin{minipage}{0.48 \linewidth}
        \centering
        \includegraphics[width=0.95\linewidth]{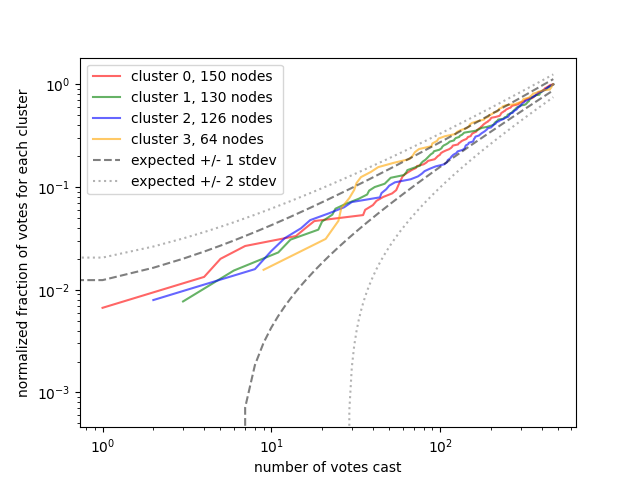}
        \setlength{\abovecaptionskip}{0pt}
        \caption{Progression of votes from each cluster for PB election 249.}
    \end{minipage}
\end{figure*}

\end{document}